\documentclass[sn-vancouver]{sn-jnl}

\usepackage{IEEEtrantools}
\usepackage{lineno}
\nolinenumbers
\usepackage{multirow}
\usepackage{color}
\usepackage{ulem}
\usepackage{amsmath,amsfonts,amssymb,bm}
\usepackage{IEEEtrantools}
\usepackage{float}
\usepackage{graphicx,color}
\usepackage{epstopdf}
\usepackage{epsfig,subfig}
\usepackage[english]{babel}
\usepackage{multirow}
\usepackage{geometry}
\newgeometry{hmargin={12mm,17mm}}   

\DeclareMathOperator*{\argmin}{arg\,min}

\jyear{2022}%

\theoremstyle{thmstyleone}%
\theoremstyle{thmstyletwo}%

\theoremstyle{thmstylethree}%

\begin{document}

\title[]{Continuous Hyper-parameter OPtimization (CHOP) in an ensemble Kalman filter}


\author*[1]{\fnm{Xiaodong} \sur{Luo}}\email{xluo@norceresearch.no}

\author[2]{\fnm{Chuan-an} \sur{Xia}}\email{c.a.xia@cugb.edu.cn}


\affil[1]{\orgname{Norwegian Research Centre (NORCE)}, \orgaddress{\city{Bergen}, \country{Norway}}}

\affil[2]{\orgname{Fuzhou University}, \orgaddress{\city{Fuzhou}, \state{Fujian Province}, \country{China}}}



\abstract{Practical data assimilation algorithms often contain hyper-parameters, which may arise due to, for instance, the use of certain auxiliary techniques like covariance inflation and localization in an ensemble Kalman filter, the re-parameterization of certain quantities such as model and/or observation error covariance matrices, and so on. Given the richness of the established assimilation algorithms, and the abundance of the approaches through which hyper-parameters are introduced to the assimilation algorithms, one may ask whether it is possible to develop a sound and generic method to efficiently choose various types of  (sometimes high-dimensional) hyper-parameters. This work aims to explore a feasible, although likely partial, answer to this question. Our main idea is built upon the notion that a data assimilation algorithm with hyper-parameters can be considered as a parametric mapping that links a set of quantities of interest (e.g., model state variables and/or parameters) to a corresponding set of predicted observations in the observation space. As such, the choice of hyper-parameters can be recast as a parameter estimation problem, in which our objective is to tune the hyper-parameters in such a way that the resulted predicted observations can match the real observations to a good extent. From this perspective, we propose a hyper-parameter estimation workflow and investigate the performance of this workflow in an ensemble Kalman filter. In a series of experiments, we observe that the proposed workflow works efficiently even in the presence of a relatively large amount (up to $10^3$) of hyper-parameters, and exhibits reasonably good and consistent performance under various conditions.}




\maketitle

\section{Introduction}\label{sec:introduction}

Data assimilation leverages the information contents of observational data to improve our understanding of quantities of interest (QoI), which could be model state variables and/or parameters, or their probability density functions (PDF) in a Bayesian estimation framework. Various challenges encountered in data assimilation problems lead to a rich list of assimilation algorithms developed from different perspectives, including, for instance, Kalman filter \citep{Kalman-new}, extended Kalman filter \citep{Simon2006}, unscented Kalman filter \citep{Julier-new}, particle filter \citep{Gordon1993,VanLeeuwen2009}, Gaussian sum filter \citep{Sorenson-recursive}, for sequential data assimilation problems; 3D- or 4-variational assimilation algorithms \citep{courtier1998ecmwf,courtier1994strategy}; and smoother algorithms for retrospective analysis \citep{cohn1994fixed}. 

To mitigate the computational costs in practical data assimilation problems, Monte Carlo or low-rank implementations of certain assimilation algorithms are investigated. Examples in this regard include ensemble Kalman filter (EnKF) and its variants \citep{Evensen-sequential,Anderson2001ensemble,Bishop-adaptive,Hamill-distance,Pham2001,hunt2007efficient,sakov2012iterative}, ensemble Kalman smoother \citep{Evensen2000}, ensemble smoother \citep{van1eeuwen996data} and their iterative versions \citep{bocquet2014iterative,emerick2012ensemble,chen2013-levenberg,luo2015Iterative}, low-rank unscented Kalman filter \citep{ambadan2009sigma,Luo-ensemble}, ensemble or low-rank Gaussian sum filter \citep{hoteit2008new,Hoteit2012,Luo2008-spgsf1}

In their practical forms, many assimilation algorithms may contain a certain number of hyper-parameters. Different from model parameters, hyper-parameters are variables that stem from assimilation algorithms and have influences on the assimilation results. As examples, one may consider the inflation factor and the length scale in covariance inflation and localization methods \citep{Anderson-Monte,Anderson2009,anderson2007exploring,bishop2011adaptive,bocquet2016localization,gharamti2018enhanced,Hamill-distance,Miyoshi2011-Gaussian,li2009simultaneous,Luo2011_EnLHF,raanes2019adaptive,zhang2004impacts}, respectively, or parameters that are related to model error simulations or representations \citep{dee1995line,dreano2017estimating,luo2019ensemble,scheffler2019inference}.   

Often, a proper choice of algorithmic hyper-parameters is essential for obtaining a descent performance of data assimilation. With the presence of various mechanisms through which algorithmic hyper-parameters are introduced, in the literature there is a vast list of methods that are proposed to estimate hyper-parameters (while sometimes relying on empirical tuning). To the best of our knowledge, it appears that the current best practice is to focus on developing tailored estimation/tuning methods for individual mechanisms. 
With this observation, a natural question would be: Is it possible to develop a common method that can be employed to estimate different types of hyper-parameters associated with an assimilation algorithm?  

This work can be considered as an attempt to find an affirmative answer to the above question. Our main idea here is to treat a data assimilation algorithm with hyper-parameters as a parametric mapping, which maps QoI (e.g., model state variables and/or parameters) to predicted observations in the observation space. From this perspective, it will be shown later that the choice of hyper-parameters can be converted to a nonlinear parameter estimation problem, which in turn can be solved through an iterative ensemble assimilation algorithm, similar to what have been done in the recent work of \cite{luo2019ensemble,scheffler2019inference}. Since ensemble-based data assimilation methods can be interpreted as some local gradient-based optimization algorithms \citep{sakov2012iterative,luo2021novel},  we impose a restriction on the hyper-parameters under estimation, that is, they have to admit continuous values. In other words, we focus on the Continuous Hyper-parameter OPtimization (CHOP) problem, whereas tuning discrete hyper-parameters is beyond the scope of the current work.

This work is organized as follows: We first formulate the CHOP problem, and propose a workflow (called CHOP workflow hereafter) to tackle the CHOP problem, which involves the use of an iterative ensemble smoother (IES) and a correlation-based adaptive localization scheme. We then investigate and report the performance of the CHOP workflow in a series of experiments. Finally, we conclude this study with some technical discussions and possible future works.  

\section{Problem statement and methodology} \label{sec:problem_statement}
\subsection{The CHOP problem} \label{subsec:chop}
We illustrate the main idea behind the CHOP workflow in the setting of a sequential data assimilation problem, in which an EnKF is adopted with a certain number of hyper-parameters. Let $\mathbf{m} \in \mathbb{R}^m$ be an $m$-dimensional vector, which contains a set of model state variables and/or parameters. In the subsequent derivation of the solution to the CHOP problem, the dynamical system is not involved. As a result, we exclude the forecast step, and focus more on the analysis step, which applies an EnKF to update a background estimation $\mathbf{m}^b$ to the analysis $\mathbf{m}^a$.       

Essentially, the EnKF can be treated as a parameterized vector mapping $\mathbf{f}_{\boldsymbol{\theta}}: \mathbf{m}^b \rightarrow \mathbf{m}^a$ that transforms $\mathbf{m}^b$ to $\mathbf{m}^a$, where $\boldsymbol{\theta}$ represents a set of algorithmic hyper-parameters to be estimated. In the context of data assimilation, the information contents of observational data, denoted by $\mathbf{d}^o \in \mathbb{R}^d$ in this work, are utilized for state and/or parameter update, whereas the update process also involves an observation operator, denoted by $\mathbf{h}$ here, which maps a background estimation $\mathbf{m}^b$ to some predicted data $\mathbf{h} \left(\mathbf{m}^b\right)$ in the observation space. 
We assume that the observations $\mathbf{d}^o$ contain some Gaussian white noise, which follows the normal distribution $N\left(\mathbf{0},\mathbf{C}_d\right)$ with mean $\mathbf{0}$ and covariance $\mathbf{C}_d$.  
In addition, we denote the background ensemble by $\mathcal{M}^b \equiv \{\mathbf{m}_j^b\}_{j=1}^{N_e}$, and the analysis ensemble by $\mathcal{M}^a \equiv \{\mathbf{m}_j^a\}_{j=1}^{N_e}$, where $j$ is the index of ensemble member, and $N_e$ represents the number of ensemble members. 

Under these settings, an analysis step of the EnKF can be represented as follows:
\begin{linenomath*}  
	\begin{IEEEeqnarray}{l} \label{eq:generic_upate_model}
		\mathbf{m}_j^a = \mathbf{f}_{\boldsymbol{\theta}} \left(\mathbf{m}_j^b,\mathcal{M}^b,\mathbf{d}_j^o, \mathbf{h}\right) \equiv  \mathbf{f} \left(\boldsymbol{\theta};\mathbf{m}_j^b,\mathcal{M}^b,\mathbf{d}_j^o, \mathbf{h}\right), \text{ for } j = 1, 2, \dotsb, N_e. 
	\end{IEEEeqnarray} 
\end{linenomath*}  
In Eq. \ref{eq:generic_upate_model}, the concrete form of the mapping $\mathbf{f}$ will depend on the specific EnKF algorithm of choice. The quantities $\mathbf{m}_j^b$, $\mathcal{M}^b$, $\mathbf{d}_j^o$ and $\mathbf{h}$ are known, whereas the hyper-parameter vector $\boldsymbol{\theta}$ is to be estimated under a certain criterion, leading to a CHOP problem.  

As an example, one may consider the case that an EnKF with perturbed observations is adopted, and covariance localization is introduced to the EnKF, such that the update formula is given as follows: 
\begin{linenomath*}  
	\begin{IEEEeqnarray}{l} \label{eq:SEnKF_with_loc}
		\mathbf{f} \left(\boldsymbol{\theta};\mathbf{m}_j^b,\mathcal{M}^b,\mathbf{d}_j^o, \mathbf{h}\right) = \mathbf{m}_j^b + \left(\mathbf{L}_{\boldsymbol{\theta}} \circ \mathbf{C}_m \right) \mathbf{h}^T \left( \mathbf{h} \left(\mathbf{L}_{\boldsymbol{\theta}} \circ \mathbf{C}_m \right) \mathbf{h}^T + \mathbf{C}_d \right)^{-1} \left(\mathbf{d}_j^o - \mathbf{h} \mathbf{m}_j^b \right).
	\end{IEEEeqnarray} 
\end{linenomath*}
In Eq. \ref{eq:SEnKF_with_loc}, we have assumed that $\mathbf{h}$ is a linear observation operator in this particular example, whereas $\mathbf{C}_m$ is the sample covariance matrix induced by the background ensemble $\mathcal{M}^b$; $\mathbf{L}_{\boldsymbol{\theta}}$ the localization matrix, which depends on some hyper-parameter(s) $\boldsymbol{\theta}$ (e.g., the length scale); and $\mathbf{L}_{\boldsymbol{\theta}} \circ \mathbf{C}_m$ stands for the Schur product of $\mathbf{L}_{\boldsymbol{\theta}}$ and $\mathbf{C}_m$. One insight from Eq. \ref{eq:SEnKF_with_loc} is that even $\mathbf{f}$ is a linear function of $\mathbf{m}_j^b$, in general $\mathbf{f}$ may have a nonlinear relation to the hyper-parameters $\boldsymbol{\theta}$.          

\subsection{Solution to the CHOP problem} \label{subsec:solution_to_chop}

In the current work, we treat CHOP as a parameter estimation problem, which can be solved through an ensemble-based, iterative assimilation algorithm, given the presence of nonlinearity in the CHOP problem. Specifically, we follow the idea in \cite{luo2015Iterative} to tackle the CHOP problem by minimizing the average of an ensemble of $N_e$ cost functions $C_j^i\left( \boldsymbol{\theta}_j^{i} \right)$ at each iteration step (indexed by $i$):
{\small
	\begin{linenomath*}  
		\begin{IEEEeqnarray}{l} 
			\label{eq:mac}\argmin_{\{\boldsymbol{\theta}_j^{i}\}_{j=1}^{N_e}} \, \dfrac{1}{N_e} \, \sum_{j=1}^{N_e} C_j^i\left( \boldsymbol{\theta}_j^{i} \right) , \\
			\label{eq:NLS} C_j^i\left( \boldsymbol{\theta}_j^{i} \right) \equiv \dfrac{1}{2}\left\{ \left(\mathbf{d}_j^o - \mathbf{g}\left( \boldsymbol{\theta}_j^{i} \right)\right)^T \mathbf{C}_{\mathbf{d}}^{-1} \left(\mathbf{d}_j^o - \mathbf{g}\left( \boldsymbol{\theta}_j^{i} \right)\right) + \gamma^{i-1}  \left( \boldsymbol{\theta}_j^{i} - \boldsymbol{\theta}_j^{i-1} \right)^T \left(\mathbf{C}_{\boldsymbol{\theta}}^{i-1}\right)^{-1} \left( \boldsymbol{\theta}_j^{i} - \boldsymbol{\theta}_j^{i-1} \right)  \right\}, \\
			\label{eq:effective_obs_sysm} \mathbf{g}\left( \boldsymbol{\theta}_j^{i} \right) \equiv  \mathbf{h} \left(\mathbf{m}_j^{i}\right) = \mathbf{h} \left(\mathbf{f} \left(\boldsymbol{\theta}_j^{i};\mathbf{m}_j^{b},\mathcal{M}^{b},\mathbf{d}_j^o, \mathbf{h}\right) \right).  
		\end{IEEEeqnarray}
	\end{linenomath*}  
}       
In Eq. \ref{eq:effective_obs_sysm}, $\mathbf{g}\left( \boldsymbol{\theta}_j^{i} \right)$, equal to $\mathbf{h} \left(\mathbf{m}_j^{i}\right)$, corresponds to the predicted observations of $\mathbf{m}_j^{i}$, which in turn depends on the hyper-parameters $\boldsymbol{\theta}_j^{i}$ for a chosen assimilation algorithm $\mathbf{f}$. 
At the end of the iteration process, suppose that in total $K$ iteration steps are executed to obtain $\boldsymbol{\theta}_j^{K}$, then we take $\mathbf{m}_j^{a} = \mathbf{m}_j^{K} = \mathbf{f} \left(\boldsymbol{\theta}_j^{K};\mathbf{m}_j^{b},\mathcal{M}^{b},\mathbf{d}_j^o, \mathbf{h}\right), \; \forall j = 1, 2, \dotsb, N_e$.    

As implied in Eqs. \ref{eq:mac} and \ref{eq:NLS}, the main idea behind the proposed CHOP workflow is to find, at each iteration step, an ensemble of hyper-parameters $\boldsymbol{\Theta}^{i} \equiv \left\{ \boldsymbol{\theta}_j^{i} \right\}_{j=1}^{N_e}$ that renders lower average data mismatch, in terms of 
\[
\sum_{j=1}^{N_e} \left(\mathbf{d}_j^o - \mathbf{g}\left( \boldsymbol{\theta}_j^{i} \right)\right)^T \mathbf{C}_{\mathbf{d}}^{-1} \left(\mathbf{d}_j^o - \mathbf{g}\left( \boldsymbol{\theta}_j^{i} \right)\right) / N_e,
\]
than the previous ensemble $\boldsymbol{\Theta}^{i-1}$ does. However, as in many ill-posed inverse problems, it is desirable to avoid over-fitting the observations. To this end, a regularization term, in the form of $\left( \boldsymbol{\theta}_j^{i} - \boldsymbol{\theta}_j^{i-1} \right)^T \left(\mathbf{C}_{\boldsymbol{\theta}}^{i-1}\right)^{-1} \left( \boldsymbol{\theta}_j^{i} - \boldsymbol{\theta}_j^{i-1} \right)$, is introduced into the cost function $C_j^i\left( \boldsymbol{\theta}_j^{i} \right)$ in Eq. \ref{eq:NLS}, whereas $\mathbf{C}_{\boldsymbol{\theta}}^{i-1}$ corresponds to the sample covariance matrix induced by the ensemble of hyper-parameters $\boldsymbol{\Theta}^{i-1} = \left\{ \boldsymbol{\theta}_j^{i-1} \right\}_{j=1}^{N_e}$, and can be expressed as  $\mathbf{C}_{\boldsymbol{\theta}}^{i-1}  = \mathbf{S}_{\theta}^{i-1} (\mathbf{S}_{\theta}^{i-1})^T $, with $\mathbf{S}_{\theta}^{i-1}$ being a square root matrix defined in Eq. \ref{eq:model_sqrt} later. The positive scalar   $\gamma^{i-1}$ can be considered a coefficient that determines the relative weight between the data mismatch and the regularization terms at each iteration step, and we will discuss its choice later.       

Another implication from Eqs. \ref{eq:mac} -- \ref{eq:effective_obs_sysm} is that instead of rendering a single estimation of the hyper-parameters, we provide an ensemble of such estimates, and each of them (e.g., $\boldsymbol{\theta}_j^{i}$) is associated with a model state and/or parameter vector $\mathbf{m}_j^{i}$. The presence of multiple estimates $\boldsymbol{\theta}_j^{i}$ not only provides the possibility of uncertainty analysis in a CHOP problem, but also avoids the need to explicitly evaluate the gradients of $\mathbf{g}$ with respect to $\boldsymbol{\theta}_j^{i}$ in the course of solving the minimization problem in Eq. \ref{eq:mac}.

Eq. \ref{eq:mac} can be approximately solved by an IES, given as follows \citep{luo2015Iterative}:
\begin{linenomath*}  
	\begin{IEEEeqnarray}{l} 
		\label{eq:mac_solution}	\boldsymbol{\theta}_j^{i} = \boldsymbol{\theta}_j^{i-1} + \mathbf{K}^{i-1} \left(\mathbf{d}_j^o - \mathbf{g}\left( \boldsymbol{\theta}_j^{i-1} \right) \right)	, \, j = 1, 2, \dotsb, N_e ; \\
		\label{eq:kalman_gain_original_IES}	\mathbf{K}^{i-1} \equiv \mathbf{S}_{\theta}^{i-1} (\mathbf{S}_{\mathbf{g}}^{i-1})^T \left(\mathbf{S}_{\mathbf{g}}^{i-1}(\mathbf{S}_{\mathbf{g}}^{i-1})^T + \gamma^{i-1} \mathbf{C}_d \right)^{-1}; \\	
		\label{eq:mean_prediction}	\bar{\boldsymbol{\theta}}^{i-1} \equiv \dfrac{1}{N_e} \sum_{j=1}^{N_e} \boldsymbol{\theta}_j^{i-1} \, ; \\
		\label{eq:model_sqrt} \mathbf{S}_{\theta}^{i-1} \equiv \dfrac{1}{\sqrt{N_e -1}} \left[\boldsymbol{\theta}_1^{i-1} - \bar{\boldsymbol{\theta}}^{i-1}, \boldsymbol{\theta}_2^{i-1} - \bar{\boldsymbol{\theta}}^{i-1}, \dotsb, \boldsymbol{\theta}_{N_e}^{i-1} - \bar{\boldsymbol{\theta}}^{i-1} \right] ; \\
		\label{eq:data_sqrt} \mathbf{S}_{\mathbf{g}}^{i-1} \equiv \dfrac{1}{\sqrt{N_e -1}} \left[\mathbf{g}\left( \boldsymbol{\theta}_1^{i-1}  \right) - \mathbf{g}\left( \bar{\boldsymbol{\theta}}^{i-1} \right), \mathbf{g}\left( \boldsymbol{\theta}_2^{i-1}  \right) - \mathbf{g}\left( \bar{\boldsymbol{\theta}}^{i-1} \right), \dotsb, \mathbf{g}\left( \boldsymbol{\theta}_{N_e}^{i-1}  \right) - \mathbf{g}\left( \bar{\boldsymbol{\theta}}^{i-1} \right) \right]  . 
	\end{IEEEeqnarray}
\end{linenomath*}                        
As one of the attractive properties of various ensemble-based assimilation algorithms, this iteration process does not explicitly involve the gradients of $\mathbf{g}$, $\mathbf{h}$ (the observation operator) or $\mathbf{f}$ (the assimilation algorithm) with respect to the hyper-parameters $\boldsymbol{\theta}$, which helps to reduce the complexities of implementing the IES algorithm.

In a practical implementation, the update formulas from Eqs. \ref{eq:mac_solution} and \ref{eq:kalman_gain_original_IES} are re-written as follows:
\begin{linenomath*}  
	\begin{IEEEeqnarray}{l} 
		\label{eq:mac_solution_normalized}	\boldsymbol{\theta}_j^{i} = \boldsymbol{\theta}_j^{i-1} + \mathbf{S}_{\theta}^{i-1} (\tilde{\mathbf{S}}_{\mathbf{g}}^{i-1})^T \left(\tilde{\mathbf{S}}_{\mathbf{g}}^{i-1}(\tilde{\mathbf{S}}_{\mathbf{g}}^{i-1})^T + \gamma^{i-1} \mathbf{I}_d \right)^{-1} \left(\tilde{\mathbf{d}}_j^o - \tilde{\mathbf{g}}\left( \boldsymbol{\theta}_j^{i-1} \right) \right); \\
		\label{eq:normalized_data_sqrt}\tilde{\mathbf{S}}_{\mathbf{g}}^{i-1} \equiv \mathbf{C}_d^{-1/2}{\mathbf{S}}_{\mathbf{g}}^{i-1}; \; \tilde{\mathbf{d}}_j^o \equiv  \mathbf{C}_d^{-1/2} \mathbf{d}_j^o; \;  \tilde{\mathbf{g}}\left( \boldsymbol{\theta}_j^{i-1} \right) \equiv \mathbf{C}_d^{-1/2} \mathbf{g}\left( \boldsymbol{\theta}_j^{i-1} \right). 	
	\end{IEEEeqnarray}
\end{linenomath*}      
In Eq. \ref{eq:mac_solution_normalized}, $\mathbf{I}_d$ represents the $d$-dimensional identity matrix. In Eq. \ref{eq:normalized_data_sqrt}, the quantities ${\mathbf{S}}_{\mathbf{g}}^{i-1}$, $\mathbf{d}_j^o$ and $\mathbf{g}\left( \boldsymbol{\theta}_j^{i-1} \right)$ in the observation space are normalized by a square root $\mathbf{C}_d^{-1/2}$ of the observation error covariance matrix. After this normalization, a singular value decomposition (SVD) is applied to $\tilde{\mathbf{S}}_{\mathbf{g}}^{i-1}$, while avoiding the potential issue of different magnitudes of observations when forming the square root matrix ${\mathbf{S}}_{\mathbf{g}}^{i-1}$. Suppose that through the SVD, we have
\begin{linenomath*}  
	\begin{IEEEeqnarray}{l} 
		\label{eq:svd} \tilde{\mathbf{S}}_{\mathbf{g}}^{i-1} = \tilde{\mathbf{U}}^{i-1} \tilde{\boldsymbol{\Sigma}}^{i-1} \left(\tilde{\mathbf{V}}^{i-1}\right)^T.	 	
	\end{IEEEeqnarray}
\end{linenomath*}      
To strengthen the numerical stability of the IES algorithm, we discard a number of relatively small singular values, which results in a truncated SVD such that
\begin{linenomath*}  
	\begin{IEEEeqnarray}{l} 
		\label{eq:tsvd} \tilde{\mathbf{S}}_{\mathbf{g}}^{i-1} \approx \hat{\mathbf{U}}^{i-1} \hat{\boldsymbol{\Sigma}}^{i-1} \left(\hat{\mathbf{V}}^{i-1}\right)^T.	 	
	\end{IEEEeqnarray}
\end{linenomath*}
The truncation criterion adopted in the current work is as follows: Suppose that the matrix $\tilde{\boldsymbol{\Sigma}}^{i-1}$ contains a number of $R$ singular values $\tilde{\sigma}_1^{i-1}, \tilde{\sigma}_2^{i-1}, \dotsb, \tilde{\sigma}_R^{i-1}$ arranged in the descending order, then we keep the first $r$ leading singular values such that $\sum_{\ell = 1}^{r} \tilde{\sigma}_{\ell}^{i-1} / \sum_{\ell = 1}^{R} \tilde{\sigma}_{\ell}^{i-1} \leq 99\%$ and $\sum_{\ell = 1}^{r+1} \tilde{\sigma}_{\ell}^{i-1} / \sum_{\ell = 1}^{R} \tilde{\sigma}_{\ell}^{i-1} > 99\%$. In Eq. \ref{eq:tsvd}, the matrix $\hat{\boldsymbol{\Sigma}}^{i-1}$ takes the leading singular values $\tilde{\sigma}_1^{i-1}, \tilde{\sigma}_2^{i-1}, \dotsb, \tilde{\sigma}_r^{i-1}$ as its diagonal elements. Accordingly, the matrices $\hat{\mathbf{U}}^{i-1}$ and $\hat{\mathbf{V}}^{i-1}$ consist of eigen-vectors that correspond to these kept leading singular values.   

Inserting Eq. \ref{eq:tsvd} into Eq. \ref{eq:mac_solution_normalized}, one obtains a modified update formula:	
\begin{linenomath*}  
	\begin{IEEEeqnarray}{l} 
		\label{eq:mac_solution_normalized_tsvd} \boldsymbol{\theta}_j^{i} \approx \boldsymbol{\theta}_j^{i-1} + \mathbf{S}_{\theta}^{i-1} \hat{\mathbf{V}}^{i-1} \hat{\boldsymbol{\Sigma}}^{i-1} \left( \left(\hat{\boldsymbol{\Sigma}}^{i-1}\right)^2 + \gamma^{i-1} \mathbf{I}_r \right)^{-1} \left(\hat{\mathbf{U}}^{i-1}\right)^T \left(\tilde{\mathbf{d}}_j^o - \tilde{\mathbf{g}}\left( \boldsymbol{\theta}_j^{i-1} \right) \right), 	 	
	\end{IEEEeqnarray}
\end{linenomath*}
which is used in all numerical experiments later. In Eq. \ref{eq:mac_solution_normalized_tsvd},  $\left(\hat{\boldsymbol{\Sigma}}^{i-1}\right)^2 \equiv \hat{\boldsymbol{\Sigma}}^{i-1} \hat{\boldsymbol{\Sigma}}^{i-1}$, and $\mathbf{I}_r$ stands for the $r$-dimensional identity matrix.  

As mentioned previously, $\gamma^{i-1}$ can be considered as a coefficient that determines the relative weight between the data mismatch and regularization terms. In the update formula, e.g., Eq. \ref{eq:mac_solution_normalized} or  \ref{eq:mac_solution_normalized_tsvd}, one can see that in effect, $\gamma^{i-1}$ affects the change $\boldsymbol{\theta}_j^{i} - \boldsymbol{\theta}_j^{i-1}$ of the hyper-parameters, which is also referred to as the step size of the iteration hereafter. Following the discussions in \cite{luo2015Iterative,luo2021novel}, it can be shown that the update formula, Eq. \ref{eq:mac_solution_normalized} or  \ref{eq:mac_solution_normalized_tsvd}, is derived by implicitly linearizing $\mathbf{g} \left(\boldsymbol{\theta}_j^{i}\right), \forall j = 1, 2, \dotsb, N_e$, around the ensemble mean $\bar{\boldsymbol{\theta}}^{i-1}$ (through the first-order Taylor approximation) at each iteration step\footnote{By ``implicitly linearizing'' we mean that the derivation of the update formula adopts the concept of linearization, but there is no need to actually evaluate the gradients of $\mathbf{g}$ with respect to $\bar{\boldsymbol{\theta}}^{i-1}$.}. In this regard, an implication is that the step size cannot be too big in order to make the linearization strategy approximately valid. On the other hand, a too small step size will slow down the convergence of the iteration process. As a result, in our implementation of the IES algorithm, e.g., Eq. \ref{eq:mac_solution_normalized}, we choose $\gamma^{i-1}$ in such a way that the influences of the two terms, $\tilde{\mathbf{S}}_{\mathbf{g}}^{i-1}(\tilde{\mathbf{S}}_{\mathbf{g}}^{i-1})^T$ and $\gamma^{i-1} \mathbf{I}_d$ are comparable (in contrast to the choice that one term dominates the other). Here, the influence is measured in terms of the trace of the respective term. As a consequence of this notion, we have $\gamma^{i-1} = \alpha^{i-1} \operatorname{trace}\left( \tilde{\mathbf{S}}_{\mathbf{g}}^{i-1}(\tilde{\mathbf{S}}_{\mathbf{g}}^{i-1})^T\right) / \operatorname{trace} \left(\mathbf{I}_d\right) = \alpha^{i-1} \operatorname{trace}\left( \tilde{\mathbf{S}}_{\mathbf{g}}^{i-1}(\tilde{\mathbf{S}}_{\mathbf{g}}^{i-1})^T\right) / d$,
where $\alpha^{i-1} > 0$ is the actual coefficient to be tuned. 

When the truncated SVD is applied to $\tilde{\mathbf{S}}_{\mathbf{g}}^{i-1}$, the choice of $\gamma^{i-1}$ for Eq. \ref{eq:mac_solution_normalized_tsvd} boils down to 
\begin{linenomath*}  
	\begin{IEEEeqnarray}{ll} 
		\label{eq:gamma_tsvd} \gamma^{i-1} & = \alpha^{i-1} \operatorname{trace}\left(\left(\hat{\boldsymbol{\Sigma}}^{i-1}\right)^2\right) / \operatorname{trace} \left(\mathbf{I}_r\right) \nonumber \\
		& = \alpha^{i-1} \sum_{\ell=1}^{r} \left(\tilde{\sigma}_{\ell}^{i-1}\right)^2 / r , 
	\end{IEEEeqnarray}
\end{linenomath*}        
At the beginning of the iteration, we let $\alpha^{0} = 1$. Subsequently, We use a backtrack line search strategy similar to that in \cite{chen2013-levenberg} to tune the coefficient value. Specifically, if the average data mismatch at step $i$ is lower than that at step $(i-1)$, then we accept the estimated hyper-parameters $\boldsymbol{\theta}_j^{i}$, and move to the next iteration step. To this end, we reduce the coefficient value by setting $\alpha^{i} = 0.9 \alpha^{i-1}$, which aims to help increase the step size at the next iteration step, similar to the idea behind the trust-region algorithm \citep{Nocedal-numerical}. 

On the other hand, if the average data mismatch value at step $i$ becomes higher than that at step $(i-1)$, then the estimated hyper-parameters $\boldsymbol{\theta}_j^{i}$ are not used for the next iteration step. Instead, a few attempts (say $K_{trial}$) are conducted to search for better estimations, leading to a so-called inner-loop iteration (if any), which is adopted for a distinction from the upper-level iteration process (called outer-loop iteration). These are done by doubling the coefficient value $\alpha_{s}^{i-1} = 2 \alpha_{s-1}^{i-1}, s = 1, 2, \dotsb, K_{trial}$,  for each trial, with $\alpha_{0}^{i-1} = \alpha^{i-1}$, and then re-running the update formula Eq. \ref{eq:mac_solution_normalized_tsvd} with a new $\gamma^{i-1}$ value calculated by Eq. \ref{eq:gamma_tsvd}, wherein the modified $\alpha_{s}^{i-1}$ value is adopted for the calculation. This strategy is again similar to the setting of the trust-region algorithm, and is also in line with the analysis in \cite{luo2015Iterative}, where it is shown that as long as the linearization strategy is approximately valid, the data mismatch values tend to decrease over the iteration steps. As such, it is sensible to increase the coefficient value (hence shrink the step size), as this helps to improve the accuracy of the first-order Taylor approximation (hence the validity of the linearization strategy). The trial process will be terminated if an average data mismatch value (obtained by using an enlarged coefficient value $\alpha_{s}^{i-1}$) is found lower than that at the iteration step $(i-1)$, or if the maximum trial number (set to $5$) is reached. At the end of the trial process, we set $\alpha^{i} = \alpha_{K_{trial}}^{i-1}$, and take $\boldsymbol{\theta}_j^{i}$ as those obtained from the last trial step.    

An additional aspect of the IES algorithm is the stopping criteria. Three such criteria are adopted in the outer-loop iteration process, which include: (1) the maximum iteration step, which is set to be $10$; (2) the threshold for the relative change of the average data mismatch values at two consecutive iteration steps, which is set to be $0.01\%$; (3) the threshold for the average data mismatch value, which is set to be $4 \times \#(\mathbf{d}^o)$ (four times the number of observations, with $\#(\mathbf{d}^o)$ being the number of elements in $\mathbf{d}^o$). In other words, the iteration process will stop if the maximum iteration step is reached. Additionally, the iteration process will also stop if the relative change of the average data mismatch values at two consecutive iteration steps, or the average data mismatch value itself at a certain iteration step, is less than their respective threshold value.

In terms of computational cost, the original analysis scheme, e.g., Eq. \ref{eq:generic_upate_model}, applies the update formula only once. In contrast, in a CHOP problem, one needs to apply the update formula multiple times during the iteration process. As such, it becomes computationally more expensive to solve the CHOP problem than a straightforward application of the EnKF analysis scheme (if one ignores the potential cost of searching for proper hyper-parameter values). In practical problems, however, the computationally most expensive part of an assimilation workflow  often lies in running the dynamical system (i.e., at the forecast step), whereas it is computationally much cheaper to execute the analysis step. Within this context, it is expected that solving the CHOP problem will only lead to a negligible (hence affordable) overhead of computational cost to the whole assimilation workflow.  

\subsection{Localization in the CHOP problem} \label{subsec:loc_for_chop}
In many data assimilation problems, the heavy cost of running the dynamical system also puts a constraint on how many ensemble members one can afford to use. Often, a trade-off has to be made so that one employs an ensemble data assimilation algorithm with a relatively small ensemble size for runtime reduction. One consequence of this limited ensemble size is that there could be substantial sampling errors when using the statistics (e.g., covariance and correlation) estimated from the small ensemble in the update formula. In addition, rank deficiencies of estimated covariance matrices would also take place. These noticed issues often lead to degraded performance of data assimilation. To mitigate the impacts of sampling errors and rank deficiency, localization techniques, e.g., \cite{anderson2007exploring,bishop2011adaptive,bocquet2016localization,Hamill-distance,janjic2011domain,fertig2007assimilating}, are often employed. 

In the CHOP problem, we note that localization is conducted with respect to hyper-parameters (e.g., in Eq. \ref{eq:mac_solution_normalized} or \ref{eq:mac_solution_normalized_tsvd}), in spite of the possible presence of another localization scheme adopted in the assimilation algorithm (e.g., as in Eq. \ref{eq:SEnKF_with_loc}). 

Many localization methods are based on the distances between the physical locations of certain pairs of quantities, which can be either pairs of two model variables as in model-space localization schemes (e.g., \citealp{Hamill-distance}), or pairs of one model variable and one observation as in observation-space localization schemes (e.g., \citealp{fertig2007assimilating}). In the CHOP problem, however, in certain circumstances it may be challenging to apply distance-based localization, as in the update formula, Eq. \ref{eq:mac_solution_normalized} or \ref{eq:mac_solution_normalized_tsvd}, certain hyper-parameters may not possess clearly defined physical locations, so that the concept of physical distance itself may not be valid.  

To circumvent this difficulty, we adopt a correlation-based adaptive localization scheme proposed in \cite{luo2019automatic}. For illustration, without loss of generality, suppose that when localization is not adopted, the update formula is in the form of      
\begin{linenomath*}  
	\begin{IEEEeqnarray}{l} 
		\label{eq:mac_solution_general}	\boldsymbol{\theta}_j^{i} = \boldsymbol{\theta}_j^{i-1} + \tilde{\mathbf{K}}^{i-1} \left(\tilde{\mathbf{d}}_j^o - \tilde{\mathbf{g}}\left( \boldsymbol{\theta}_j^{i-1} \right) \right), 	
	\end{IEEEeqnarray}
\end{linenomath*}  
where $\tilde{\mathbf{K}}^{i-1}$ is a Kalman-gain-like matrix and $\left(\tilde{\mathbf{d}}_j^o - \tilde{\mathbf{g}}\left( \boldsymbol{\theta}_j^{i-1} \right) \right)$ the corresponding innovation term. With the presence of localization, then the update formula is modified as
\begin{linenomath*}  
	\begin{IEEEeqnarray}{l} 
		\label{eq:mac_solution_general_with_loc}	\boldsymbol{\theta}_j^{i} = \boldsymbol{\theta}_j^{i-1} + \left(\mathbf{L}^{i-1} \circ \tilde{\mathbf{K}}^{i-1}\right) \left(\tilde{\mathbf{d}}_j^o - \tilde{\mathbf{g}}\left( \boldsymbol{\theta}_j^{i-1} \right) \right), 	
	\end{IEEEeqnarray}
\end{linenomath*}  
where $\mathbf{L}^{i-1}$ is a $h \times d$ localization matrix to be constructed, with $h$ and $d$ being the vector lengths of $\boldsymbol{\theta}_j^{i}$ and $\tilde{\mathbf{g}}\left( \boldsymbol{\theta}_j^{i-1} \right)$ (or $\tilde{\mathbf{d}}_j^o$), respectively. In Eq. \ref{eq:mac_solution_general_with_loc}, the localization scheme is similar to observation-space localization, but the localization matrix $\mathbf{L}^{i-1}$ acts on the Kalman-gain-like matrix $\tilde{\mathbf{K}}^{i-1}$.     

The construction of the localization matrix $\mathbf{L}^{i-1}$ is based on the notion of causality detection between the hyper-parameters $\boldsymbol{\theta}_j^{i}$ and the predicted observations $\tilde{\mathbf{g}}\left( \boldsymbol{\theta}_j^{i-1} \right)$ \citep{luo2019automatic}. To see the rationale behind this notion, let $\tilde{\mathbf{d}}_j^{i-1,pred} \equiv \tilde{\mathbf{g}}\left( \boldsymbol{\theta}_j^{i-1} \right)$ and $\Delta \tilde{\mathbf{d}}_j^{i-1} \equiv \tilde{\mathbf{d}}_j^o - \tilde{\mathbf{d}}_j^{i-1,pred}$, and re-write Eq. \ref{eq:mac_solution_general_with_loc} into an equivalent, element-wise form     
\begin{linenomath*}  
	\begin{IEEEeqnarray}{l} 
		\label{eq:mac_solution_general_with_loc_scalar}	\theta_{j,s}^{i} = \theta_{j,s}^{i-1} + \sum_{t=1}^{d} \left(L_{s,t}^{i-1} \tilde{K}_{s,t}^{i-1}\right) \Delta \tilde{d}_{j,t}^{i-1}, \text{ for } s = 1, 2, \dotsb, h, 
	\end{IEEEeqnarray}
\end{linenomath*} 
where $\theta_{j,s}^{i}$, $\theta_{j,s}^{i-1}$ and $\Delta \tilde{d}_{j,t}^{i-1}$ represent the $s-$th or the $t-$th element of $\boldsymbol{\theta}_j^{i}$, $\boldsymbol{\theta}_j^{i-1}$ and $\Delta \tilde{\mathbf{d}}_j^{i-1}$, respectively; while $L_{s,t}^{i-1} \in \left[0,1\right]$ and $\tilde{K}_{s,t}^{i-1}$ stand for the elements on the $s-$th row and the $t-$th column of the matrices $\mathbf{L}^{i-1}$ and $\tilde{\mathbf{K}}^{i-1}$, respectively.

The implication of Eq. \ref{eq:mac_solution_general_with_loc_scalar} is that the innovation elements $\Delta \tilde{d}_{j,t}^{i-1}$ ($t = 1, 2, \dotsb, d$) contribute to the change $\theta_{j,s}^{i} - \theta_{j,s}^{i-1}$ of the $s-$th hyper-parameter, and the degree of the contribution of each innovation element $\Delta \tilde{d}_{j,t}^{i-1}$ is determined by the element $\tilde{K}_{s,t}^{i-1}$ (if no localization), together with the tapering coefficient $L_{s,t}^{i-1}$ (if with localization).      

In the notion of causality detection to choose the value of $L_{s,t}^{i-1}$, the main idea is that if there is a causality from the $s-$th element of hyper-parameters to the $t-$th element of innovations, then $\Delta \tilde{d}_{j,t}^{i-1}$ should be used for updating $\theta_{j,s}^{i-1}$ to $\theta_{j,s}^{i}$, meaning that $L_{s,t}^{i-1} \neq 0$. In contrast, if there is no causality therein, then it is sensible to exclude $\Delta \tilde{d}_{j,t}^{i-1}$ so that it makes no contribution to the update of $\theta_{j,s}^{i-1}$ to $\theta_{j,s}^{i}$, meaning that $L_{s,t}^{i-1} = 0$.            

Here, the statistics used to measure the causality is the sample cross correlations (e.g., denoted by $\rho_{s,t}^{i-1}$) between the elements of an ensemble of hyper-parameters (e.g., $\theta_{j,s}^{i-1}$ for $j = 1, 2, \dotsb, N_e$) and the corresponding ensemble of innovations (e.g., $\Delta \tilde{d}_{j,t}^{i-1}$  for $j = 1, 2, \dotsb, N_e$). Intuitively, when the magnitude of a sample correlation, say $\rho_{s,t}^{i-1}$,  is relatively high (e.g., close to $1$), then one tends to believe that there is a true causality from the $s-$th element of hyper-parameters to the $t-$th element of innovations. On the other hand, when the magnitude of $\rho_{s,t}^{i-1}$ is relatively low (e.g., close, but not exactly equal, to $0$), then more caution is needed. This is because when a limited ensemble size $N_e$ is adopted, the induced sampling errors can cause spurious correlations, such that even there is no causality between a hyper-parameter and an innovation element, the estimated sample correlation may not be identical to zero.   

Taking into account the above consideration, we assign values to $L_{s,t}^{i-1}$ following a method in \cite{luo2019automatic}:
\begin{linenomath*}  
	\begin{IEEEeqnarray}{l} 
		\label{eq:loc_rule}	L_{s,t}^{i-1} =  f_{GC} \left( \dfrac{1-\vert \rho_{s,t}^{i-1} \vert}{1- 3/\sqrt{N_e}} \right), N_e > 9, 
	\end{IEEEeqnarray}
\end{linenomath*}  
where $f_{GC}$ is the Gaspari-Cohn (GC) function \citep{Gaspari1999}, which, for a scalar input $z \geq 0$,  satisfies  
\begin{linenomath*}  
	\begin{equation} \label{ch2_cov_filtering_correlation_func}
		f_{GC} \left( z \right) = \begin{cases}
			-\dfrac{1}{4} z^5 + \dfrac{1}{2} z^3 + \dfrac{5}{8} z^3 - \dfrac{5}{3} z^2 +1 \, , & \text{if}~ 0 \le z \le 1 \, ; \\
			\dfrac{1}{12} z^5 - \dfrac{1}{2} z^4 + \dfrac{5}{8} z^3 + \dfrac{5}{3} z^2 - 5 z + 4 - \dfrac{2}{3} z^{-1} \, , & \text{if}~ 1 < z \le 2 \, ;\\
			0 \, , & \text{if}~ z >2 \, . \\ 
		\end{cases} 
	\end{equation}
\end{linenomath*} 

In Eq. \ref{eq:loc_rule}, the factor $3/\sqrt{N_e}$ is adopted for the following reason: When the true correlation between the $s$-th hyper-parameter and the $t$-th innovation is $0$, but the sample correlation is evaluated with a sample size of $N_e$, then the sampling errors follow a Gaussian distribution $N(0,1/N_e)$ asymptotically, see \cite{luo2019automatic} and the reference therein. Therefore, under the hypothesis (denoted by $H_0$ hereafter) that the true correlation is $0$, we compare the magnitude of the sample correlation $\rho_{s,t}^{i-1}$ with three times the standard deviation (STD) ($3/\sqrt{N_e}$). The larger $\vert \rho_{s,t}^{i-1} \vert$ is, the more confident we are that $H_0$ should be rejected, meaning it is more likely that there is a true (non-zero) correlation between the $s$-th hyper-parameter and the $t$-th innovation. As such, $L_{s,t}^{i-1}$ will receive a larger value. On the other hand, the value of $L_{s,t}^{i-1}$ becomes smaller as $\vert \rho_{s,t}^{i-1} \vert$ decreases.

In comparison to distance-based localization, a few additional benefits of the above correlation-based localization include: better abilities to hand non-local observations, time-lapse effects of observations and big observation datasets; and improved adaptivity to different types of model parameters/state variables. For more details, readers are referred to \cite{luo2019automatic}.

\section{Numerical results} 
The L96 model \citep{Lorenz-optimal} is taken as the testbed in the current study. For a $N_{L}$-dimensional L96 model, its dynamic behavior is described by the following ordinary differential equations (ODEs): 
\begin{linenomath*}
	\begin{equation} \label{eq:LE98}
		\frac{dx_e}{dt} = \left( x_{e+1} - x_{e-2} \right) x_{e-1} - x_e + F, \, e=1, \dotsb, N_L.
	\end{equation}
\end{linenomath*}
For consistency, $x_{-1}=x_{N_L -1}$, $x_{0}=x_{N_L}$ and $x_{1}=x_{N_L + 1}$ in Eq.~(\ref{eq:LE98}). The driving force term $F$ is set to $8$ throughout this work. The L96 model is integrated forward in time by the fourth-order Runge-Kutta method with a constant integration step of $0.05$ time units (dimensionless).

In the experiments, a few statistics are adopted to characterize the performance of data assimilation. These include the root mean square error (RMSE) $E_m$, ensemble spread $S_{en}$ and data mismatch $E_d$. As will be seen below, RMSE computes a normalized euclidean distance between an estimate and the ground truth in the model space, whereas data mismatch calculates a similar distance between predicted and real observations in the observation space. On the other hand, ensemble spread provides a measure of ensemble variability.  

To compute these statistics, let $\mathbf{m}$ be a $m$-dimensional vector of estimated model state variables and/ or parameters that are of interest, $\mathbf{d}^{pred} \equiv \mathbf{h} \left( \mathbf{m} \right)$ the corresponding predicted observation, with $\mathbf{h}$ being the observation operator, then given the reference $\mathbf{m}^{ref}$ (ground truth), we define the RMSE of $\mathbf{m}$ as
\begin{linenomath*}
	\begin{equation} \label{eq:rmse}
		E_m = \Vert \mathbf{m} - \mathbf{m}^{ref}  \Vert_2 / \sqrt{m} , 
	\end{equation}
\end{linenomath*}  
where the operator $\Vert \bullet \Vert_2$ returns the euclidean norm of its operand $\bullet$.   

In addition, assume that the real observation is $\mathbf{d}^{o}$, which is contaminated by some zero-mean Gaussian white noise, and is associated with an observation error covariance matrix $\mathbf{C}_d$, then we define the data mismatch of $\mathbf{m}$ as
\begin{linenomath*}
	\begin{equation} \label{eq:data_mismatch}
		E_d = \left( \mathbf{d}^{o} - \mathbf{d}^{pred}\right)^T \mathbf{C}_d^{-1} \left( \mathbf{d}^{o} - \mathbf{d}^{pred}\right) . 
	\end{equation}
\end{linenomath*}         

For the definition of ensemble spread, let $\mathcal{M} = \left\{\mathbf{m}_j \equiv \left[m_{j,1}, m_{j,2}, \dotsb m_{j,m}\right]^T \right\}_{j=1}^{N_e}$ be an ensemble of estimated model state variables/parameters, where $m_{j,k}$ denotes the $k$-th element of $\mathbf{m}_j$ ($k = 1, 2, \dotsb, m$). Based on $\mathcal{M}$, we construct a vector $\mathbf{S} \equiv \left[\sigma_1, \sigma_2, \dotsb, \sigma_m \right]^T$, where $\sigma_k$ denotes the sample standard deviation with respect to the ensemble $\{m_{j,k}\}_{j=1}^{N_e}$, and compute the ensemble spread as 
\begin{linenomath*}
	\begin{equation} \label{eq:ensemble_spread}
		S_{en} = \Vert \mathbf{S}  \Vert_2 / \sqrt{m} . 
	\end{equation}
\end{linenomath*}    

\subsection{Experiments in a 40-dimensional L96 system}

\subsubsection{Experiment settings}
We start from the common choice of $N_L = 40$ in the literature, while considering a much larger $N_L$ value later on. We run the L96 model from time $0$ to time $5000$ (which corresponds to $100,000$ integration steps in total), and compute the long-term (lt) temporal mean $\hat{\mathbf{m}}^{lt}$ and covariance $\hat{\mathbf{C}}^{lt}$ based on the model variables at all integration steps.

In each of the experiments below, we draw a random sample from the Gaussian distribution $N\left(\hat{\mathbf{m}}^{lt}, \hat{\mathbf{C}}^{lt}\right)$, and use this sample as the initial condition to start the simulation of the L96 model in a transition time window of 250 time units (corresponding to $5000$ integration steps). 

The model variables obtained at the end of the transition time window is then taken as the initial values to simulate reference model variables in an assimilation time window of 250 time units. Data assimilation is conducted within this assimilation time window to estimate reference model variables at different time steps, based on a background ensemble of model variables and noisy observations that are related to reference model variables through a certain observation system. The initial background ensemble (at the first time instance of the assimilation time window) is generated by drawing a specified number $N_e$ of samples from the Gaussian distribution $N\left(\hat{\mathbf{m}}^{lt}, \hat{\mathbf{C}}^{lt}\right)$. The ensemble size $N_e$ may change with the experiments, as will be specified later. 

For a generic vector $\mathbf{m}$ of model state variables/parameters, the observation system adopted in the experiments is linear and in the form of 
\begin{linenomath*}
	\begin{equation} \label{eq:obs_system}
		\begin{split}
			\mathbf{d} & = \mathbf{H} \mathbf{m} \\ 
			& = \left[m_1,m_{1 + \Delta n}, m_{1 + 2\Delta n}, \dotsb, m_{ 1 + M \Delta n} \right]^T , 
		\end{split}
	\end{equation}
\end{linenomath*} 
where $\mathbf{H}$ is a matrix extracting elements $m_1,m_{1 + \Delta n}, m_{1 + 2\Delta n}, \dotsb$ from $\mathbf{m}$,  the integer $\Delta n$ represents an increment of model-variable index, and $M$ is the largest integer such that $ 1 +  M \Delta n \leq N_L$. The value of $\Delta n$ may also vary in different experiments. As such, its concrete value will be mentioned in individual experiments later. For convenience, hereafter we may also use the shorthand notation $\{1:\Delta n : N_L\}$ to denote the set $\{1, 1 + \Delta n, 1 + 2\Delta n, \dotsb\}$ of indices. Similar notations will also be used elsewhere later.     

In the experiments, we assume that the observation operator $\mathbf{H}$ is perfect and known to us. When applying Eq. \ref{eq:obs_system} to reference model variables to generate real observations for data assimilation,  we add to the outputs of Eq. \ref{eq:obs_system} some Gaussian white noise $\boldsymbol{\epsilon}$, which is assumed to follow the Gaussian distribution $N(\mathbf{0}_{M+1}, \mathbf{I}_{M+1})$, with $\mathbf{0}_{M+1}$ and $\mathbf{I}_{M+1}$ being the $(M+1)$-dimensional zero vector, and the $(M+1)$-dimensional identity matrix, respectively. The frequency for us to collect the measurements is every $\Delta t$ integration steps, whose value will also be specified in respective experiments.  

The base assimilation algorithm adopted here is the EnKF with perturbed observations \citep{Burgers-analysis}, in which the update formula reads:
\begin{linenomath*}  
	\begin{IEEEeqnarray}{l} \label{eq:SEnKF}
		\mathbf{m}_j^a = \mathbf{m}_j^b + \mathbf{C}_m \mathbf{H}^T \left( \mathbf{H} \mathbf{C}_m \mathbf{H}^T + \mathbf{C}_d \right)^{-1} \left(\mathbf{d}_j^o - \mathbf{H} \mathbf{m}_j^b \right), \text{ for } j = 1, 2, \dotsb, N_e, 
	\end{IEEEeqnarray} 
\end{linenomath*}    
where $\mathbf{C}_m$ is the sample covariance matrix of the background ensemble $\mathcal{M}^b \equiv \{\mathbf{m}_j^b\}_{j=1}^{N_e}$, and $\mathbf{d}_j^o$ stands for perturbations with respect to the real observation $\mathbf{d}^o$. 

Covariance inflation and localization are then introduced to Eq. \ref{eq:SEnKF} to strengthen the performance of the EnKF. We note that our purpose here is to demonstrate how the CHOP workflow can be implemented on top of certain chosen inflation and localization techniques, yet the CHOP workflow itself cannot be used to design new inflation or localization techniques.  

Specifically, in this study, covariance inflation is conducted on the background ensemble, in such a way that $\mathcal{M}^b$ is replaced by a modified background ensemble $\tilde{\mathcal{M}}^b \equiv \{\tilde{\mathbf{m}}_j^b\}_{j=1}^{N_e}$ with $\tilde{\mathbf{m}}_j^b = \bar{\mathbf{m}}^b + \left( 1+\delta \right) \left( \mathbf{m}_j^b -  \bar{\mathbf{m}}^b \right)$, where $\bar{\mathbf{m}}^b$ is the ensemble mean of the members in $\mathcal{M}^b$, and $\delta \geq 0$ is the inflation factor to be determined through a certain criterion. Accordingly, the sample covariance $\mathbf{C}_m$ in Eq. \ref{eq:SEnKF} should be replaced by $ \tilde{\mathbf{C}}_m = \left( 1+\delta \right)^2 \mathbf{C}_m$, which is larger than $\mathbf{C}_m$ (hence the name covariance inflation).     

On the other hand, localization is implemented by replacing the Kalman gain matrix $\tilde{\mathbf{K}} = \tilde{\mathbf{C}}_m \mathbf{H}^T \left( \mathbf{H} \tilde{\mathbf{C}}_m \mathbf{H}^T + \mathbf{C}_d \right)^{-1}$ 
by the Schur product $\mathbf{L} \circ \tilde{\mathbf{K}}$, where $\mathbf{L}$ is the localization matrix, whose element, say, $L_{s,t}$ on the $s$-th row and the $t$-th column of $\mathbf{L}$, is determined by the ``physical'' distance between the $s$-th model variable $m_s$ and the $t$-th observation element $d_t$. For the observation system in Eq. \ref{eq:obs_system}, $d_t$ corresponds to the observation at the model-variable location $o = (1 + (t-1) \Delta n)$ (in terms of model-variable index). As such, the element $L_{s,t}$ is computed as follows:
\begin{linenomath*}  
	\begin{IEEEeqnarray}{ll} 
		\label{eq:tapering_rule} L_{s,t} & = f_{GC} \left( \dfrac{ dist_{s,t} }{\lambda} \right), \\
		\label{eq:dist_rule} dist_{s,t} & = \min \left(|s - o|/N_L, 1 -  |s - o|/N_L \right) .
	\end{IEEEeqnarray} 
\end{linenomath*}    
In Eq. \ref{eq:tapering_rule}, $f_{GC}$ is the Gaspari-Cohn function (see Eq. \ref{ch2_cov_filtering_correlation_func}), $dist_{s,t}$ represents a normalized distance between the $s$-th model variable and the $t$-th observation element (which is located on the $o$-th model grid/index), and $\lambda$ is the length scale, whose value is chosen under a certain criterion. Eq. \ref{eq:dist_rule} computes the distance between the $t$-th and $o$-th model grids/indices, which is normalized by the total number $N_L$ of the model grids (equal to the dimension of the L96 model in this case). Note that $dist_{s,t}$ takes the minimum value between $|s - o|/N_L$ and $1 -  |s - o|/N_L$, due to the circular nature of the L96 model. In the sequel, we re-write $\mathbf{L}$ as $\mathbf{L}\left(\lambda\right)$ to indicate the dependence of $\mathbf{L}$ on $\lambda$.    

Taking into account the presence of both covariance inflation and localization, the base assimilation algorithm, Eq. \ref{eq:SEnKF}, is modified as follows:
{\small
	\begin{linenomath*}  
		\begin{IEEEeqnarray}{ll} \label{eq:SEnKF_inf_loc}
			\mathbf{m}_j^a =  \left[ \bar{\mathbf{m}}^b + \left( 1+\delta \right) \left( \mathbf{m}_j^b -  \bar{\mathbf{m}}^b \right) \right]  + & \left\{ \mathbf{L}\left(\lambda \right) \circ \left[ \mathbf{C}_m \mathbf{H}^T \left( \mathbf{H} \mathbf{C}_m \mathbf{H}^T + \mathbf{C}_d / \left(1 + \delta \right)^2 \right)^{-1} \right] \right\}  \left(\mathbf{d}_j^o - \mathbf{H} \left[ \bar{\mathbf{m}}^b + \left( 1+\delta \right) \left( \mathbf{m}_j^b -  \bar{\mathbf{m}}^b \right) \right] \right). 
		\end{IEEEeqnarray} 
	\end{linenomath*}
}    

The update formula in Eq. \ref{eq:SEnKF_inf_loc} thus contains two hyper-parameters, the inflation factor $\delta$ and the length scale $\lambda$. With the known background ensemble $\mathcal{M}^b$ (hence $\mathbf{m}_j^b$, $\bar{\mathbf{m}}^b$ and $\mathbf{C}_m$) and the quantities $\mathbf{d}_j^o$, $\mathbf{C}_d$ and $\mathbf{H}$, the relation between the analysis $\mathbf{m}_j^a$ and the hyper-parameters is complex (and nonlinear in general), even with a rather simple observation operator $\mathbf{H}$.   

Eq. \ref{eq:SEnKF_inf_loc} serves as the reference algorithm hereafter, and we will compare its performance with that of the CHOP workflow in a number of different experiments below. In the comparison, we do not adopt any tailored methods proposed in the literature to tune $\delta$ and/or $\lambda$. Instead, we use the grid search method to find the optimal values of the pair $(\delta_{min}, \lambda_{min})$, whereas the optimality is meant in the sense that the combination $(\delta_{min}, \lambda_{min})$ results in the lowest value of an average RMSE within some pre-defined search ranges of $\delta$ and $\lambda$. In all the experiments related to the $40$-dimensional L96 model, for the reference algorithm Eq. \ref{eq:SEnKF_inf_loc}, the search range of $\delta$ is set to $\{0:0.1:2\}$, and that of $\lambda$ to $\{0.05:0.05:1\}$. For a given experiment, the average RMSE is obtained by first computing the RMSEs of all analysis ensemble means at different time instances, then averaging these RMSEs over the whole assimilation time window, and finally averaging the previous (average) values again over a number of repetitions of the assimilation run. These repetitions share identical experimental settings, except that the random seeds used to generate certain random variables (e.g., the initial background ensemble and the observation noise) in each repetition of the experiment are different. In each experiment with respect to the $40$-dimensional L96 model, the number of repetitions is set to $20$. 

In the CHOP workflow, instead of relying on the grid search method to find an optimal combination of $\delta$ and $\lambda$, the IES algorithm presented in Section \ref{sec:problem_statement} is applied to estimate an ensemble of $\delta$ and $\lambda$ values for the reference algorithm Eq. \ref{eq:SEnKF_inf_loc}. Note that there are differences between the optimality criterion used in the grid search method and that in the CHOP workflow. In this regard, the grid search method aims to find a single optimal pair $(\delta_{min}, \lambda_{min})$ that leads to the globally minimum average RMSE in the model space, within the whole assimilation time window. In contrast, the CHOP workflow searches for an ensemble of $\delta$ and $\lambda$ values that help reduce the average of an ensemble of data mismatch values in the observation space (cf Eq. \ref{eq:mac}) within a given number of iteration steps, and at each data assimilation cycle (rather than the whole assimilation time window). In this sense, the obtained ensemble of $\delta$ and $\lambda$ values represents, at best, locally optimal estimates at a given time instance, with a prescribed maximum number of iteration steps.

With these aforementioned differences, it is natural to expect that the globally optimal criterion (global criterion for short) used in the grid search method should result in better data assimilation performance than the locally optimal one (local criterion for short) adopted in the CHOP workflow. On the other hand, it is important to notice that the superiority of the global criterion is achieved on top of the assumption that one has access to the ground truths of model state variables and/or parameters during the whole data assimilation window. As such, it is not a realistic criterion that can be applied to practical data assimilation problems, where the underlying ground truths are typically unknown. In contrast, the local criterion is more realistic and can be implemented in practice. In the experiments below, however, we still choose to present the results with respect to the global criterion, as this serves as a means to cross-validate the performance of the CHOP workflow.  

In the CHOP workflow, the configuration of the IES algorithm is as follows: Eqs. \ref{eq:mac_solution_normalized_tsvd} and \ref{eq:gamma_tsvd} are employed to estimate ensembles of hyper-parameters $\left\{\boldsymbol{\theta}_j^i \equiv \left[\delta_j^i,\lambda_j^i\right]^T \right\}_{j=1}^{N_e}$ at different iteration steps (indexed by $i$, for $i = 1, 2, ..., K$), and correlation-based localization is applied to Eq. \ref{eq:mac_solution_normalized_tsvd} (in addition to distance-based localization adopted in the reference algorithm Eq. \ref{eq:SEnKF_inf_loc}). We note that the size of a hyper-parameter ensemble is the same as that of a background ensemble $\mathcal{M}^b = \{\mathbf{m}_j^b\}_{j=1}^{N_e}$ of model state variables and/or parameters, so that each ensemble member $\mathbf{m}_j^b$ is associated with its respective hyper-parameter pair $\left(\delta_j^i,\lambda_j^i\right)$, when using the reference algorithm Eq. \ref{eq:SEnKF_inf_loc} to update $\mathbf{m}_j^b$. To start the iteration process of the CHOP workflow, Latin hypercube sampling (LHS) is adopted to generate an initial ensemble of hyper-parameters at each assimilation cycle, whereas the hyper-parameter ranges used for LHS are the same as those in the grid search method.   

Another remark is that the background ensemble $\mathcal{M}^b$ already exists before the CHOP workflow starts, and is invariant during the iteration process of the CHOP workflow. On the other hand, the outputs of the reference algorithm Eq. \ref{eq:SEnKF_inf_loc} do depend on the values of $\left(\delta_j^i,\lambda_j^i\right)$, and  can change as the iteration proceeds. The members $\mathbf{m}_j^a$ of the analysis ensemble are taken as the outputs of Eq. \ref{eq:SEnKF_inf_loc} at the last iteration step $K$, which is a number jointly determined by the three stopping criteria mentioned previously (cf Section \ref{sec:problem_statement}).   

\subsubsection{Results with different ensemble sizes}
\begin{table*} [!t]
	\centering
	\caption{\label{tab:min_rmse_full_obs} Performance comparison between the grid search method and the CHOP workflow applied to the reference algorithm Eq. \ref{eq:SEnKF_inf_loc} in the full observation scenario, with four different ensemble sizes. For the grid search method, we report the minimum average RMSEs within the search ranges, and their associated STDs. In addition, we also present the combination of the inflation factor and the length scale, $(\delta_{min},\lambda_{min})$, that results in the minimum average RMSE in each experiment. For the CHOP workflow, the inflation factor and the length scale are estimated at each assimilation cycle, and thus vary with time. As such, we only report the average RMSEs and their associated STDs.}
	\begin{tabular}{cccc}
		\hline
		\hline 
		\multirow{2}{*}{Ensemble size}  & \multicolumn{2}{c}{Grid search} & CHOP \\
		\cline{2-4}
		& Minimum average RMSE (mean $\pm$ STD) & $(\delta_{min},\lambda_{min})$ & Average RMSE (mean $\pm$ STD) \\ 
		\hline               
		$N_e = 15$ & $0.5235 \pm 0.0104$   & $(0.15,0.15)$  &  $1.2212 \pm 0.1832$    \\
		\hline  
		$N_e = 20$ & $0.4845 \pm 0.0112$  & $(0.15,0.25)$ &   $0.6180 \pm 0.0353$   \\
		\hline  
		$N_e = 25$ & $0.4711 \pm 0.0059$   & $(0.15,0.30)$ & $0.5080 \pm 0.0167$     \\
		\hline  
		$N_e = 30$ & $0.4560 \pm 0.0100$  & $(0.10,0.20)$ &  $0.4766 \pm 0.0096$     \\  
		\hline 
		\hline
	\end{tabular}
\end{table*}   
\newcommand{\nScale}{0.17}
\begin{figure} 
	\centering
	\begin{tabular}{cc}
		\subfloat[$N_e = 15$]{\includegraphics[scale=\nScale]{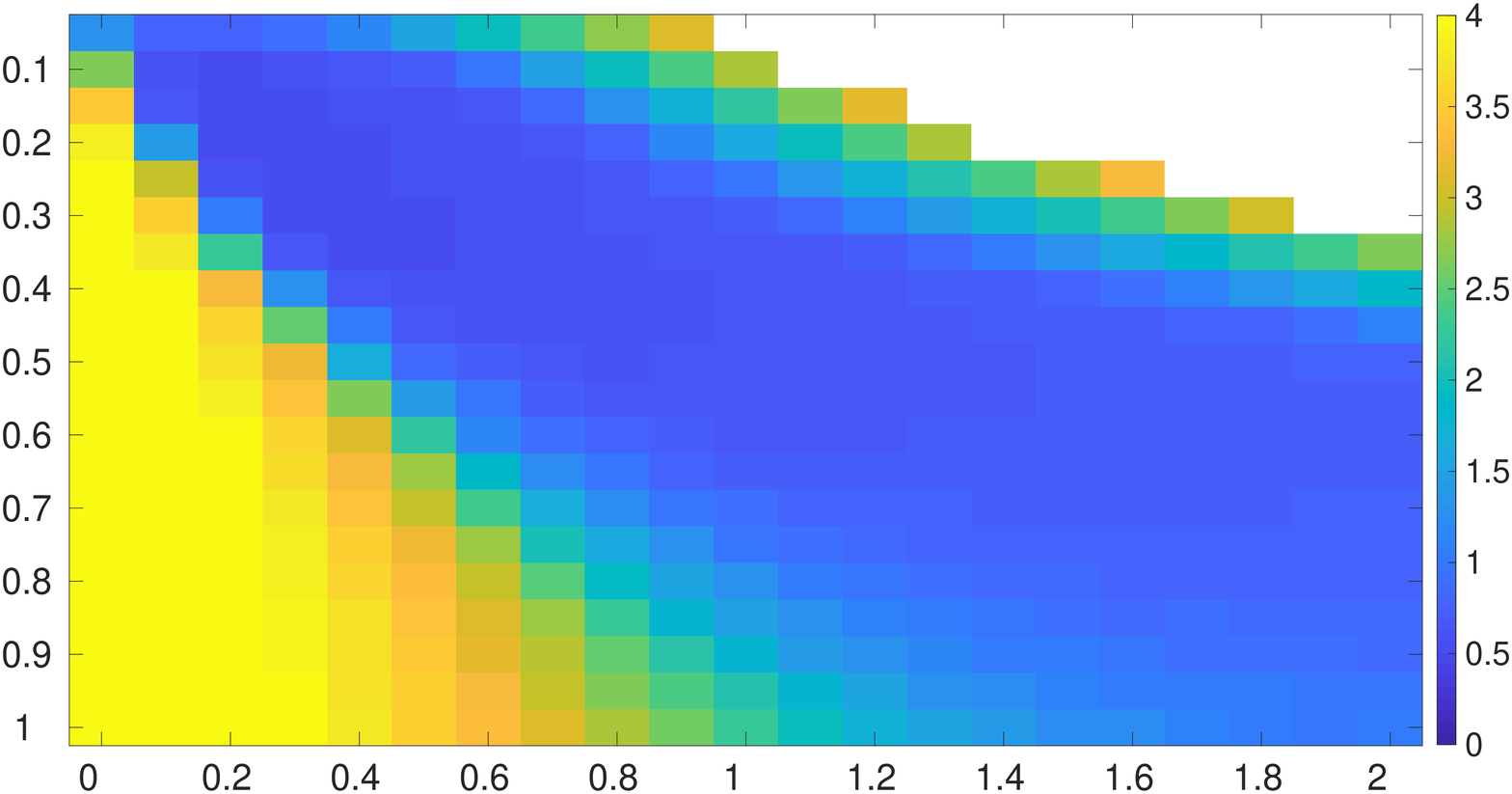}} &
		\subfloat[$N_e = 20$]{\includegraphics[scale=\nScale]{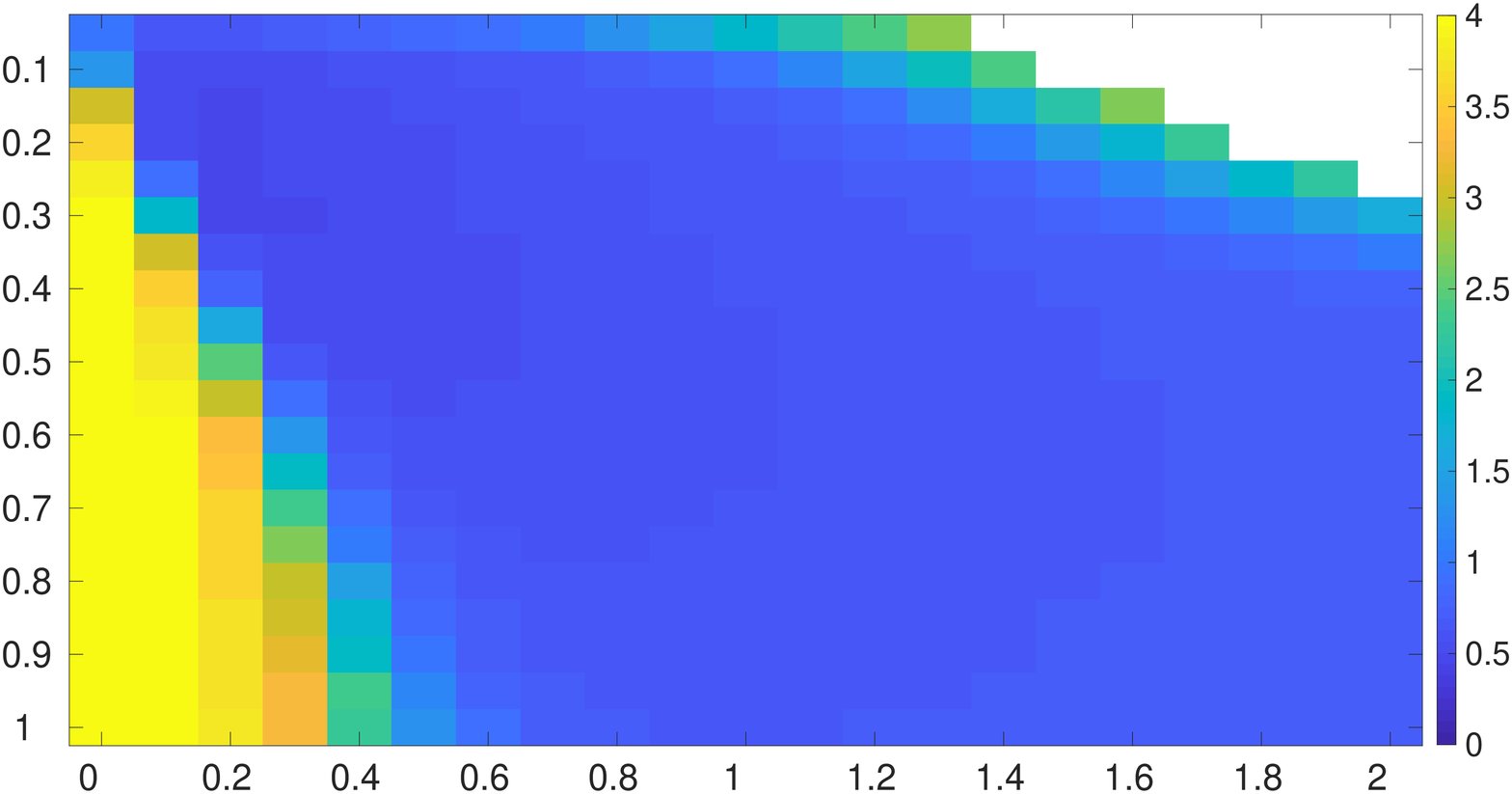}} \\ 
		\subfloat[$N_e = 25$]{\includegraphics[scale=\nScale]{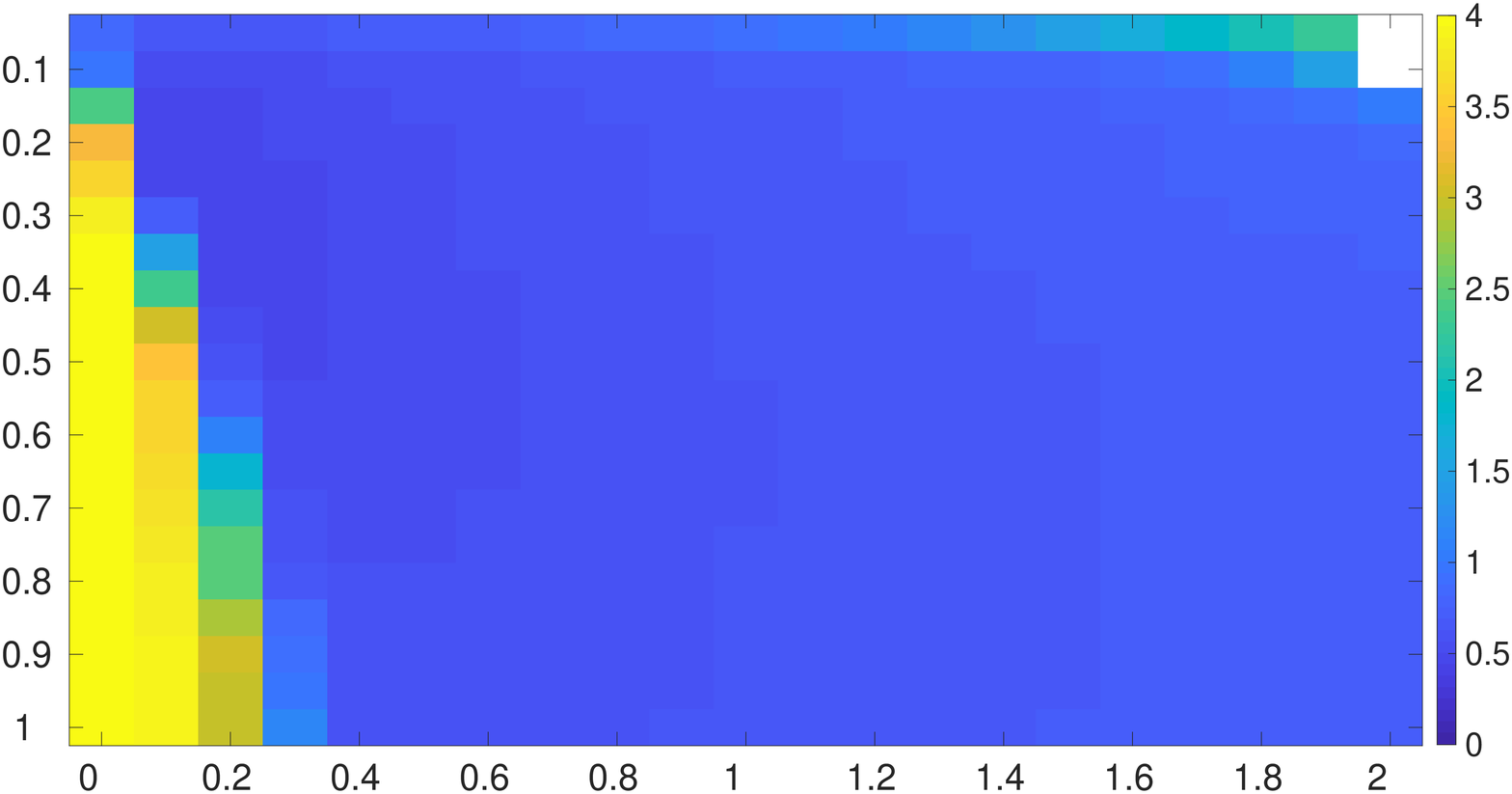}} & 
		\subfloat[$N_e = 30$]{\includegraphics[scale=\nScale]{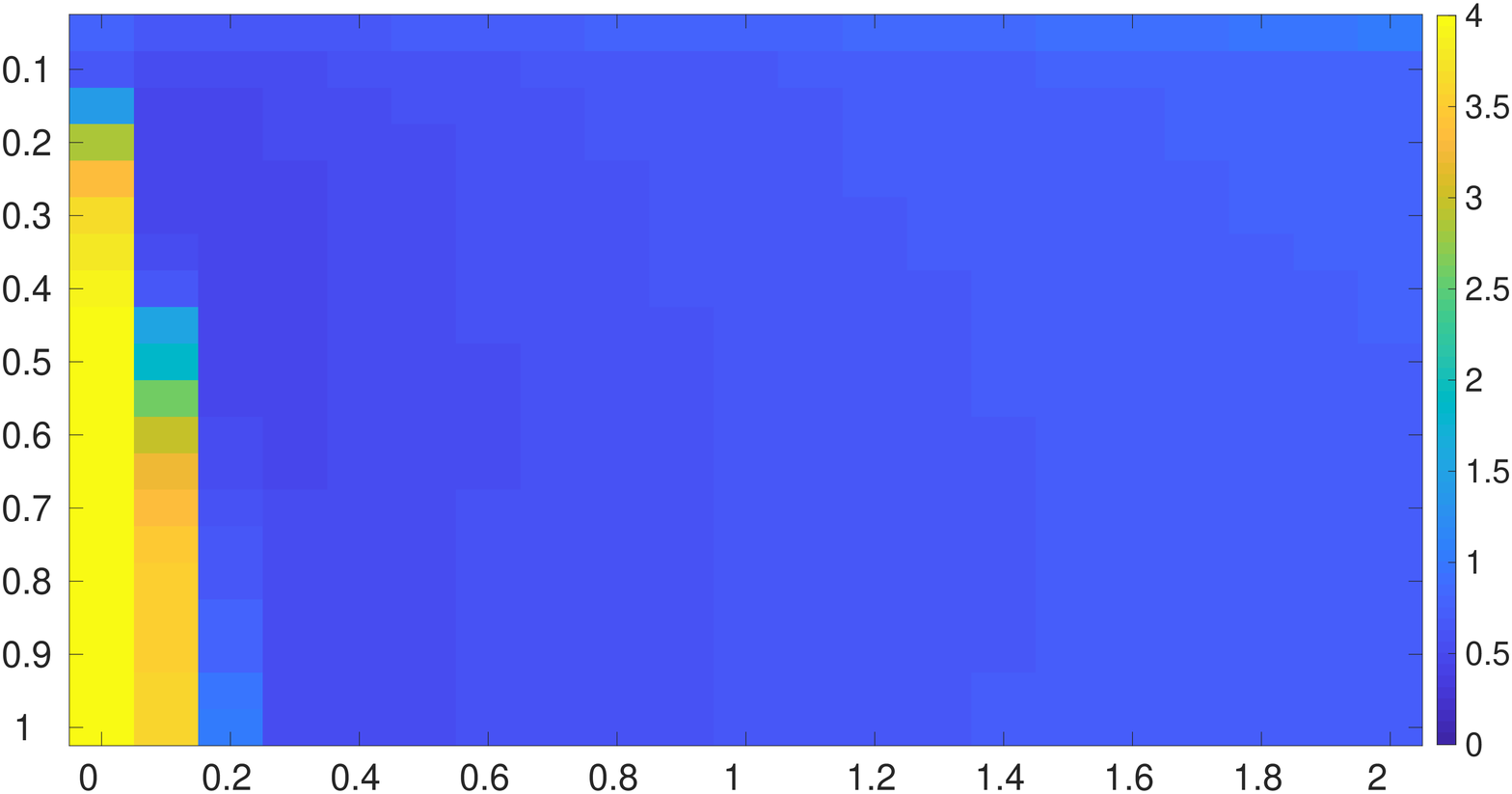}} \\ 
	\end{tabular}
	\caption{\label{fig:rmse_inf_loc_full_obs} Average RMSEs with respect to the reference algorithm Eq. \ref{eq:SEnKF_inf_loc} in the full observation scenario ($\Delta n = 1$, $N^{freq} = 4$), using an ensemble size of $15$, $20$, $25$ and $30$, respectively. The RMSE values are obtained by searching all the possible combinations of the inflation factor $\delta \in \{0:0.1:2\}$ (along the horizontal axis) and the length scale $\lambda \in \{0.05:0.05:1\}$ (along the vertical axis). Note that for certain combinations of $\delta$ and $\lambda$ values, filter divergence may take place (represented by white color in respective sub-plots).}
\end{figure} 

We first present results in a set of four experiments to illustrate the impacts of ensemble size. In each experiment, all state variables are observed (called full observation scenario hereafter), corresponding to the observation-index increment $\Delta n = 1$, with an observation frequency of every $4$ integration steps (denoted by $N^{freq} = 4$). These four experiments use ensemble sizes $N_e = 15, 20, 25, 30$, respectively, while the remaining experimental settings (e.g., real observations/perturbed observations, initial background ensemble) are identical. 

Figure \ref{fig:rmse_inf_loc_full_obs} shows the average RMSEs in the full observation scenario, obtained by applying the grid search method to the reference algorithm Eq. \ref{eq:SEnKF_inf_loc}, when different ensemble sizes $N_e$ are used in the experiments. 

For a given ensemble size, the sub-plots of Figure \ref{fig:rmse_inf_loc_full_obs} indicate that in general, relatively low average RMSEs are reached with suitable amounts of covariance inflation and localization, whereas relatively high average RMSEs are obtained if there are insufficient inflation (corresponding to relatively small $\delta$ values) and localization (corresponding to relatively large $\lambda$ values). On the other hand, too strong inflation (corresponding to relatively large $\delta$ values) and localization (corresponding to relatively small $\lambda$ values) may lead to filter divergence (represented by white color in the sub-plots)\footnote{If filter divergence takes place in any repetition run, then we assign NaN (not a number) to the average RMSE.}, which corresponds to the situation where the RMSE values blow up with an potential issue of numerical overflow.    

On the other hand, comparing the sub-plots of Figure \ref{fig:rmse_inf_loc_full_obs}, it can be observed that a larger ensemble size tends to result in a larger area that is filled with relatively low average RMSEs, while reducing the chance of filter divergence. 

In company with Figure \ref{fig:rmse_inf_loc_full_obs}, Table \ref{tab:min_rmse_full_obs} reports the minimum average RMSEs that the grid search method can achieve in the four sets of experiments, their associated STDs (to reflect the degrees of  fluctuations of the average RMSEs within 20 repetition runs), and the optimal combinations $(\delta_{min},\lambda_{min})$ of the inflation factor and the length scale, with which the minimum average RMSEs are achieved. As one can see therein, when the ensemble size increases, the minimum average RMSE obtained by the grid search method tends to decrease. Meanwhile, less amounts of covariance inflation (in the sense of smaller $\delta_{min}$) and localization (in the sense of larger $\lambda_{min}$) are required to achieve the minimum average RMSE, consistent with the observations in Figure \ref{fig:rmse_inf_loc_full_obs}.   

%
For comparison, Table \ref{tab:min_rmse_full_obs} also lists the average RMSEs that are obtained by the CHOP workflow in the full observation scenario. Note that the CHOP workflow uses the IES to estimate an ensemble of inflation factors and length scales at each assimilation cycle. As such, unlike the grid search method, there is no time-invariant, globally optimal inflation factor or length scale obtained from the CHOP workflow.               

A few observations can be obtained when comparing the performance of the grid search method and the CHOP workflow in Table \ref{tab:min_rmse_full_obs}. First of all, in terms of the minimum average RMSE that one can achieve in each experiment, the CHOP workflow systematically under-performs the grid search method. This under-performance is not surprising, since, as discussed previously, the grid search method gains the relative superiority on top of the assumption that it has access to the ground truths, which is typically infeasible in practical data assimilation problems. 

In comparison to the grid search method, the CHOP workflow appears to be more sensitive to the change of ensemble size. With $N_e = 15$, there is a relatively large gap (around $0.7$) between the average RMSE of the CHOP workflow and the minimum average RMSE that the grid search method can achieve. As the ensemble size increases, the performance of the CHOP workflow substantially improves,  such that the gap drops to only around $0.02$ when $N_e = 30$. This indicates that in the full observation scenario, the CHOP workflow can perform reasonably well with a sufficiently large ensemble size.      

\subsubsection{Results with different observation densities}  
\begin{table*} [!t]
	\centering
	\caption{\label{tab:min_rmse_half_quarter_obs} As in Table \ref{tab:min_rmse_full_obs}, but for performance comparison between the grid search method and the CHOP workflow with full, half, quarter and octantal observations, respectively, whereas the ensemble size and the observation frequency are set to $30$ and $4$, respectively, in all experiments.}
	\begin{tabular}{cccc}
		\hline
		\hline 
		\multirow{2}{*}{Observation density}  & \multicolumn{2}{c}{Grid search} & CHOP \\
		\cline{2-4}
		& Minimum average RMSE (mean $\pm$ STD) & $(\delta_{min},\lambda_{min})$ & Average RMSE (mean $\pm$ STD) \\ 
		\hline  
		Full ($\Delta n = 1$) & $0.4560 \pm 0.0100$  & $(0.10,0.20)$ &  $0.4766 \pm 0.0096$     \\
		Half ($\Delta n = 2$) & $0.7975 \pm 0.0257$  & $(0.10,0.20)$ &  $0.8763 \pm 0.0418$     \\ 
		Quarter ($\Delta n = 4$) & $2.0100 \pm 0.0773$  & $(0.10,0.25)$ &  $2.3596 \pm 0.1248$     \\  
		Octantal ($\Delta n = 8$) & $2.9129 \pm 0.0353$  & $(0.05,0.10)$ &  $3.2437 \pm 0.0419$     \\  
		\hline 
		\hline
	\end{tabular}
\end{table*}   
\renewcommand{\nScale}{0.17}
\begin{figure} 
	\centering
	\begin{tabular}{cc}
		\subfloat[Full observation scenario ($\Delta n = 1$)]{\includegraphics[scale=\nScale]{./ETKF_rmseMean_ensize30_obvSkip1_skipStep4_obsNoiseLevel1}}  &
		\subfloat[Half observation scenario ($\Delta n = 2$)]{\includegraphics[scale=\nScale]{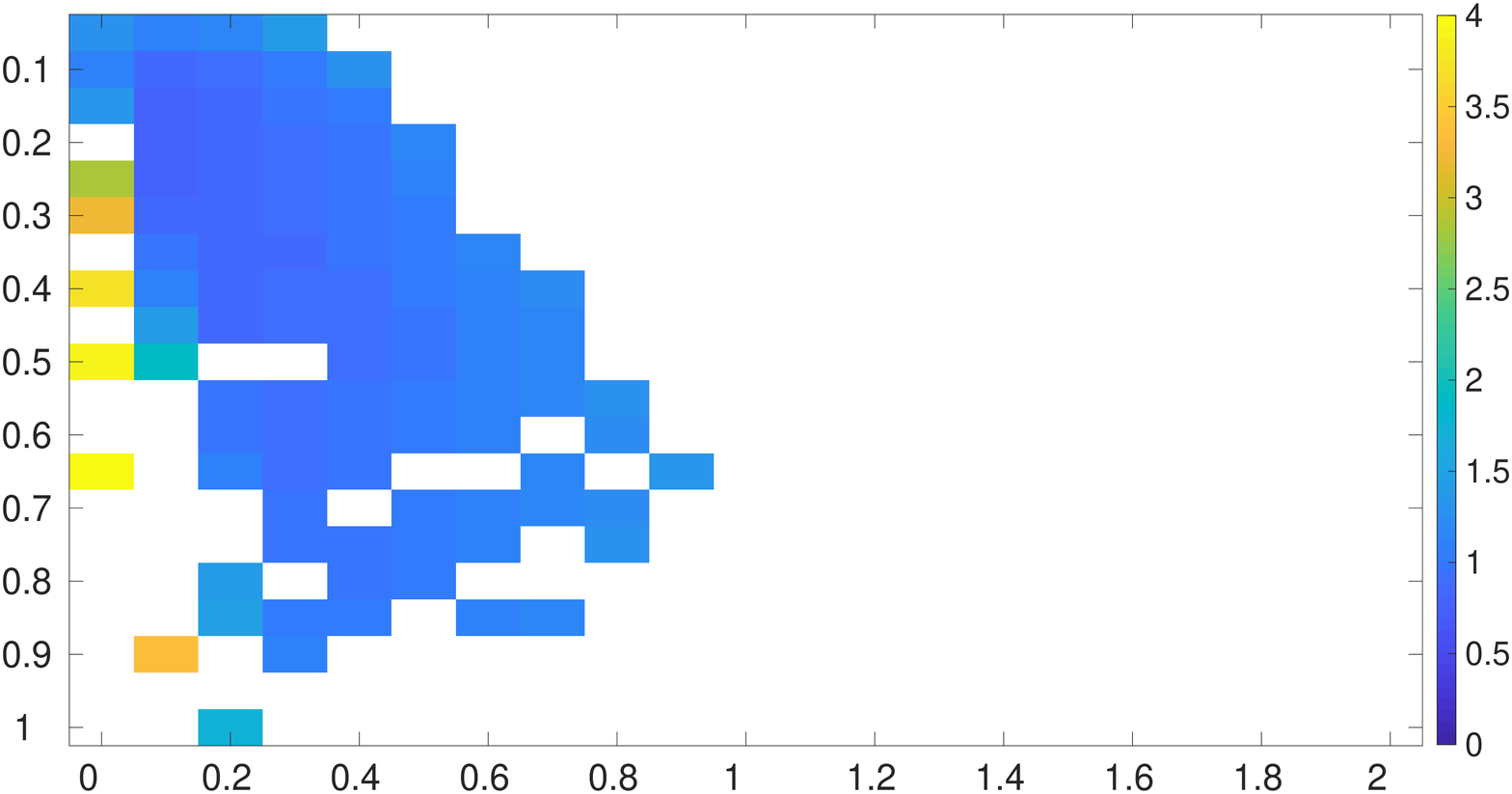}} \\
		\subfloat[Quarter observation scenario ($\Delta n = 4$)]{\includegraphics[scale=\nScale]{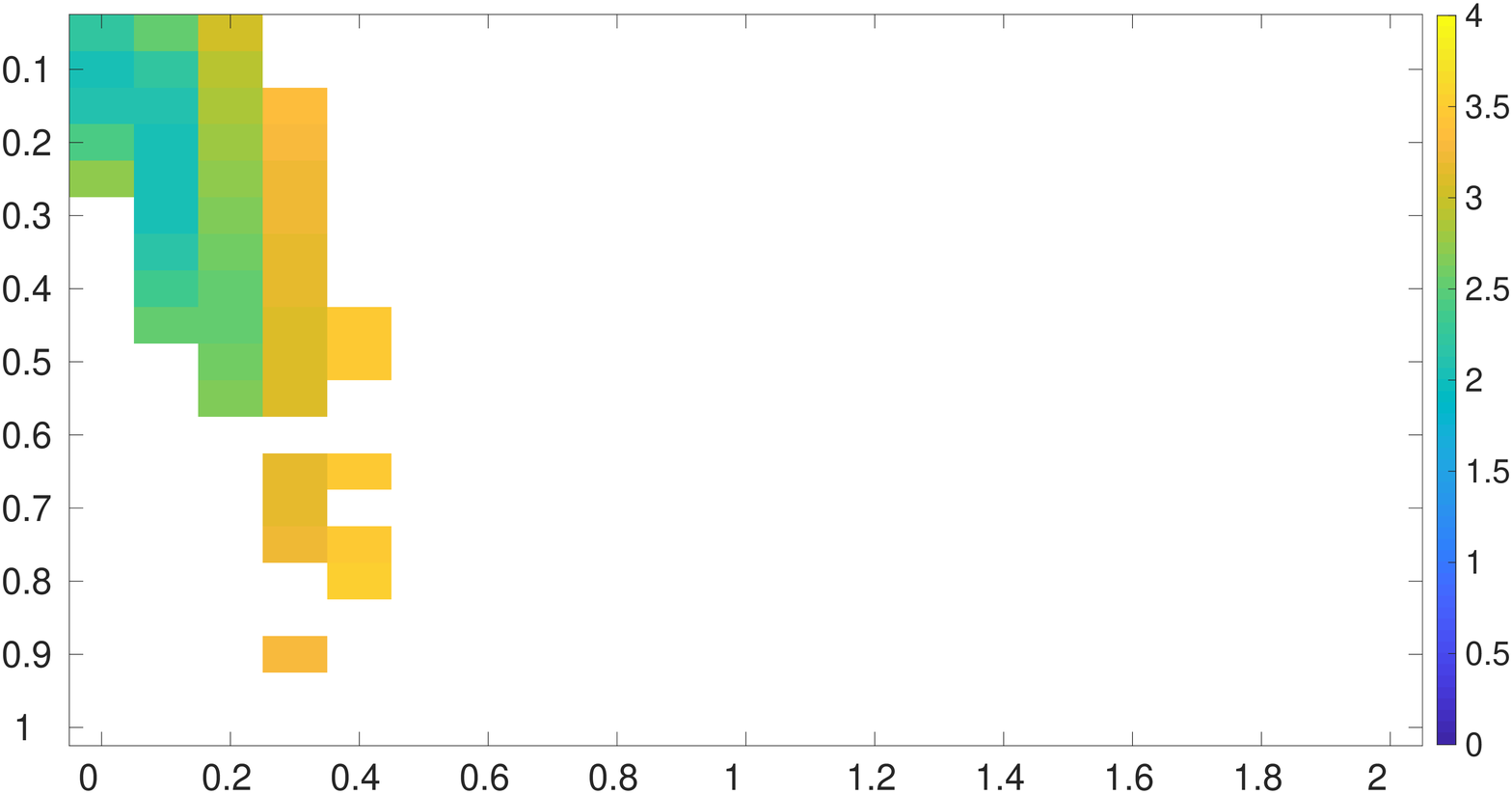}} &
		\subfloat[Octantal observation scenario ($\Delta n = 8$)]{\includegraphics[scale=\nScale]{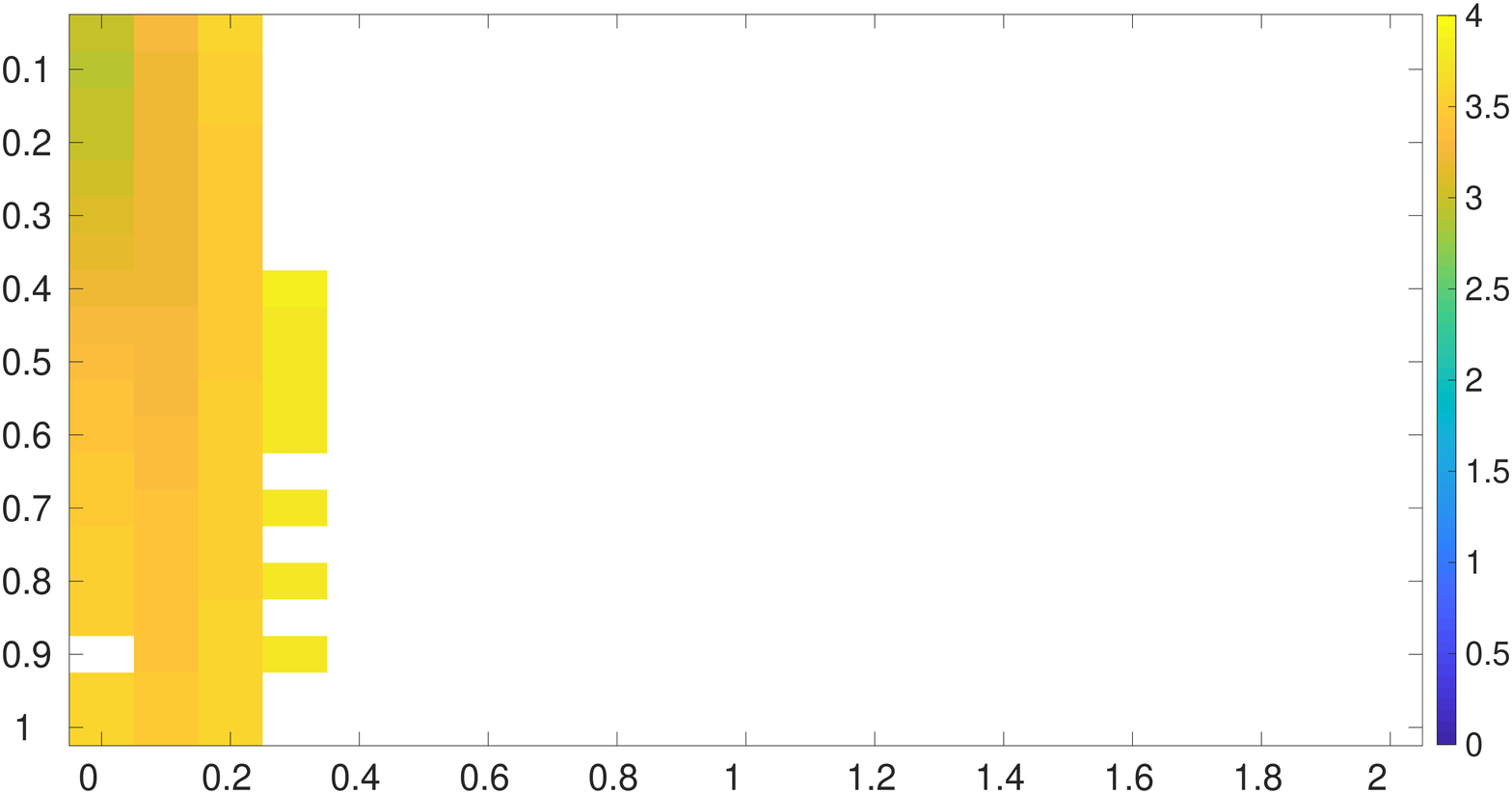}} \\ 
	\end{tabular}
	\caption{\label{fig:rmse_inf_loc_half_quarter_obs} As in Figure \ref{fig:rmse_inf_loc_full_obs}, but for average RMSEs obtained by the grid search method in the half ($\Delta n = 2$, $N^{freq} = 4$), quarter ($\Delta n = 4$, $N^{freq} = 4$) and octantal ($\Delta n = 8$, $N^{freq} = 4$)  observation scenarios, respectively, with the ensemble sizes $N_e = 30$. For ease of comparison, the results of the full observation scenario ($\Delta n = 1$, $N^{freq} = 4$, $N_e = 30$) in Figure \ref{fig:rmse_inf_loc_full_obs} are re-plotted here.}
\end{figure} 

We then examine the impact of observation density on the performance of the grid search method and the CHOP workflow. To this end, we conduct three more experiments with the observation-index increment $\Delta n = 2$ (the half observation scenario), $\Delta n = 4$ (the quarter observation scenario), $\Delta n = 8$ (the octantal observation scenario) respectively, while these three experiments share the same ensemble size $N_e = 30$ and observation frequency $N^{freq} = 4$. 

Figure \ref{fig:rmse_inf_loc_half_quarter_obs} reports the average RMSEs with different combinations of the inflation factor and length scale values, obtained by the grid search method in the half, quarter and octantal observation scenarios, respectively. For convenience of comparison, the results of the full observation scenario (with $N_e = 30$) in Figure \ref{fig:rmse_inf_loc_full_obs}(d) are re-plotted therein. Comparing the results in Figure \ref{fig:rmse_inf_loc_half_quarter_obs}, it can be seen that, as the observation density decreases ($\Delta n$ increases), the performance of the grid search method degrades, in the sense that the resulted average RMSEs arise, and filter divergence tends to have a higher chance to take place, except that the quarter observation scenario seems to have more instances of filter divergence than the octantal observation scenario. The degraded performance is expected, since reduced observation density means that less information contents can be utilized for data assimilation. 

Similar to Table \ref{tab:min_rmse_full_obs}, Table \ref{tab:min_rmse_half_quarter_obs} posts the minimum average RMSEs of the grid search method, their associated STDs, and the optimal values of the inflation factor and the length scale. Among the full, half and quarter observation scenarios, as the observation density decreases, the optimal inflation factor $\delta_{min}$ does not change, but the optimal length scale $\lambda_{min}$ shows a tendency of increment, meaning that less localization is required. This trend, however, is broken in the octantal observation scenario, in which both $\delta_{min}$ and $\lambda_{min}$ become smaller than those of the other three scenarios, suggesting that it is better to have less inflation but more localization.    

For comparison, Table \ref{tab:min_rmse_half_quarter_obs} also lists the average RMSEs with respect to the CHOP workflow. As one can see therein, in different observation scenarios, the average RMSEs of the CHOP workflow stay in a relatively close vicinity of the minimum values achieved by the grid search method. In addition, no filter divergence is spotted in the repetition runs of the CHOP workflow. As such, the CHOP workflow again appears to work reasonably well with different observation densities.  

\subsubsection{Results with different observation frequencies}       
\begin{table*} [!t]
	\centering
	\caption{\label{tab:min_rmse_frequency} As in Table \ref{tab:min_rmse_full_obs}, but for performance comparison between the grid search method and the CHOP workflow in the half observation scenario ($\Delta n = 2$), with the same ensemble size $N_e = 30$ yet different observation frequencies.}
	\begin{tabular}{cccc}
		\hline
		\hline 
		\multirow{2}{*}{Observation frequency}  & \multicolumn{2}{c}{Grid search} & CHOP \\
		\cline{2-4}
		& Minimum average RMSE (mean $\pm$ STD) & $(\delta_{min},\lambda_{min})$ & Average RMSE (mean $\pm$ STD) \\ 
		\hline  
		$N^{freq} = 1$ & $0.3948 \pm 0.0124$  & $(0.10,0.45)$ &  $0.5409 \pm 0.0117$   \\
		$N^{freq} = 2$ & $0.5015 \pm 0.0123$  & $(0.10,0.30)$ &  $0.5471 \pm 0.0193$     \\
		$N^{freq} = 4$ & $0.7975 \pm 0.0257$  & $(0.10,0.20)$ &  $0.8763 \pm 0.0418$     \\ 
		$N^{freq} = 8$ & $1.8369 \pm 0.0557$  & $(0.10,0.20)$ &  $2.1022 \pm 0.0473$     \\  
		\hline 
		\hline
	\end{tabular}
\end{table*}   
\renewcommand{\nScale}{0.17}
\begin{figure} 
	\centering
	\begin{tabular}{cc}
		\subfloat[$N^{freq} = 1$]{\includegraphics[scale=\nScale]{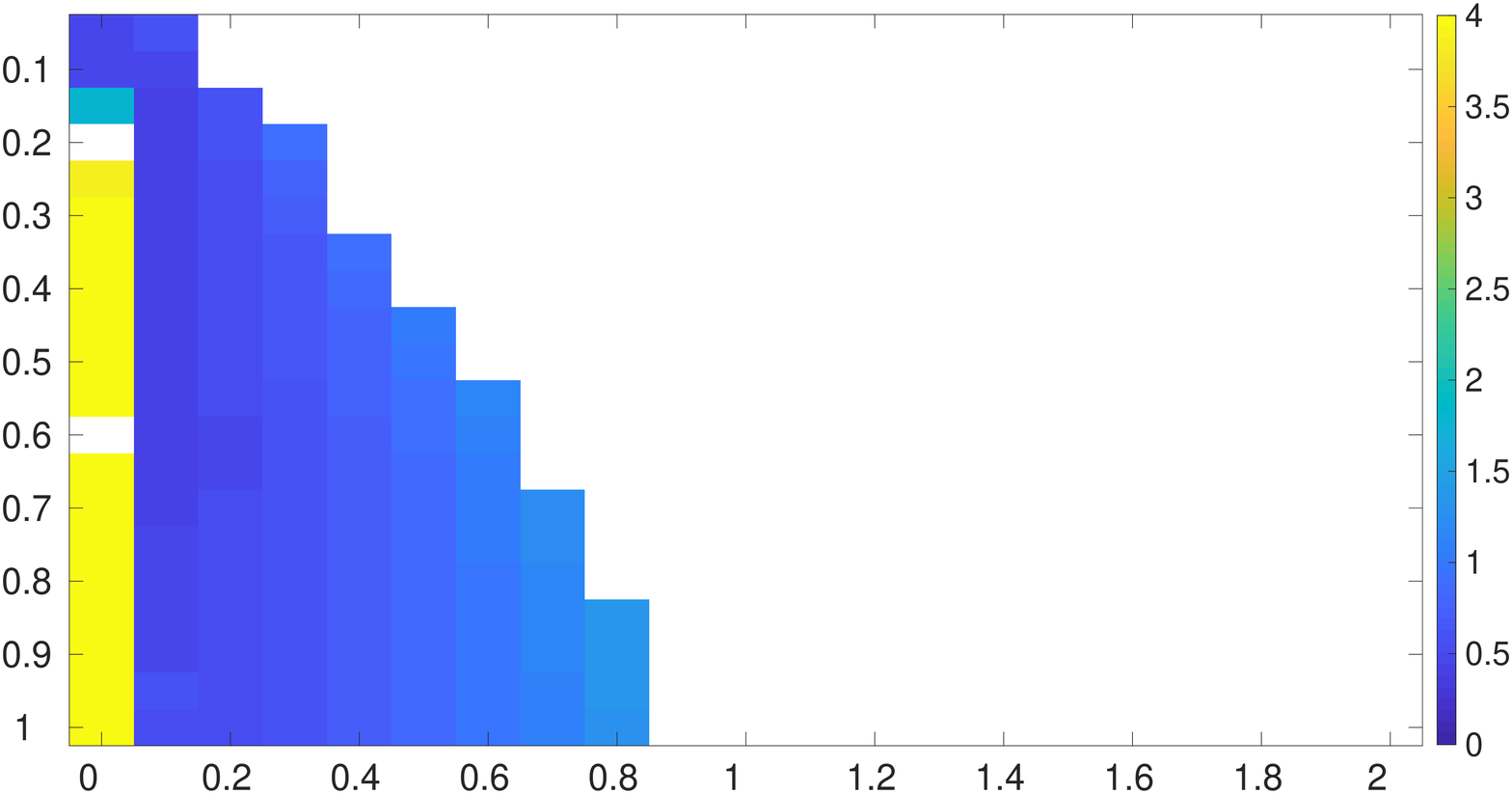}} &
		\subfloat[$N^{freq} = 2$]{\includegraphics[scale=\nScale]{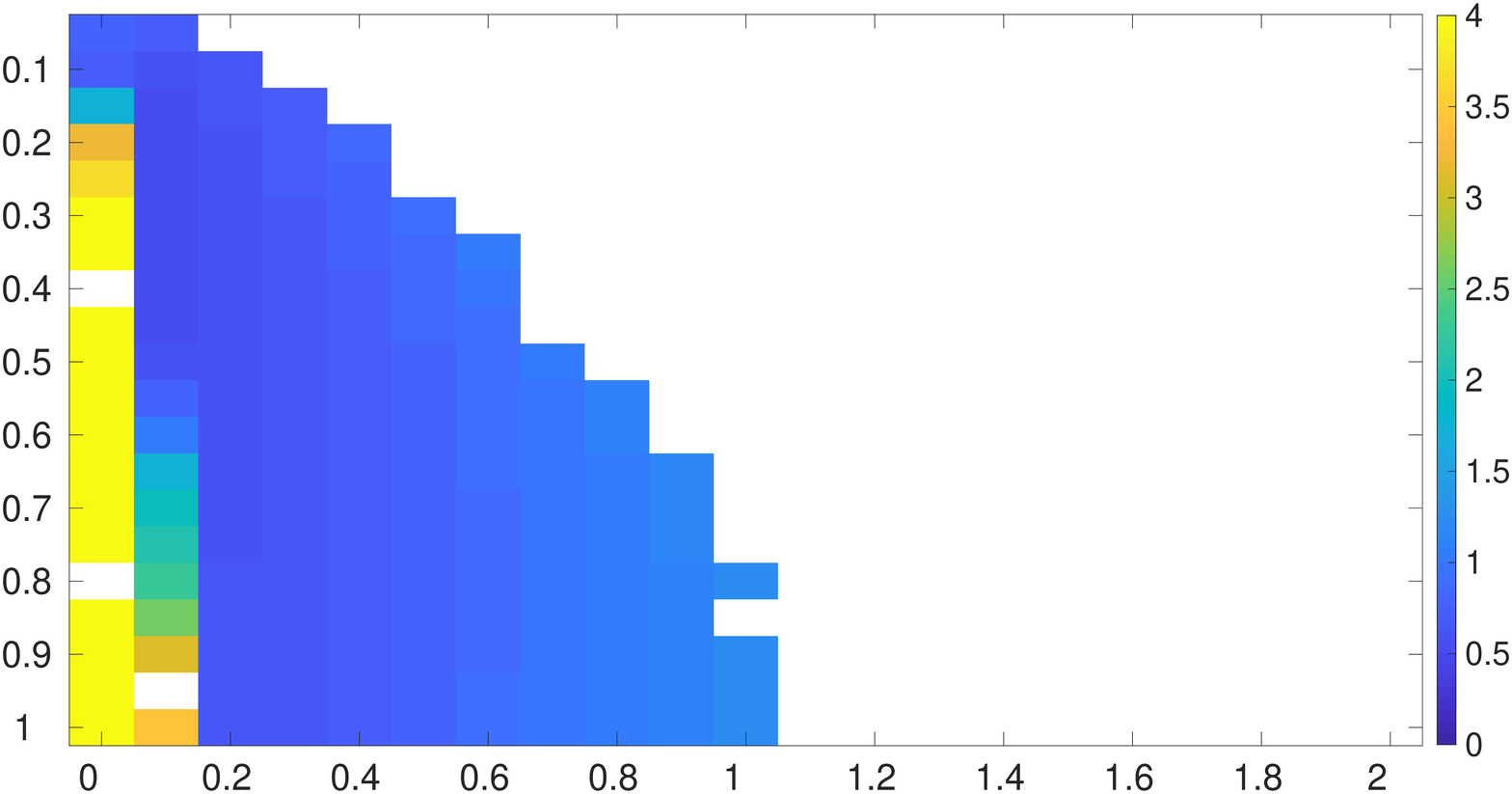}} \\ 
		\subfloat[$N^{freq} = 4$]{\includegraphics[scale=\nScale]{./ETKF_rmseMean_ensize30_obvSkip2_skipStep4_obsNoiseLevel1}} & 
		\subfloat[$N^{freq} = 8$]{\includegraphics[scale=\nScale]{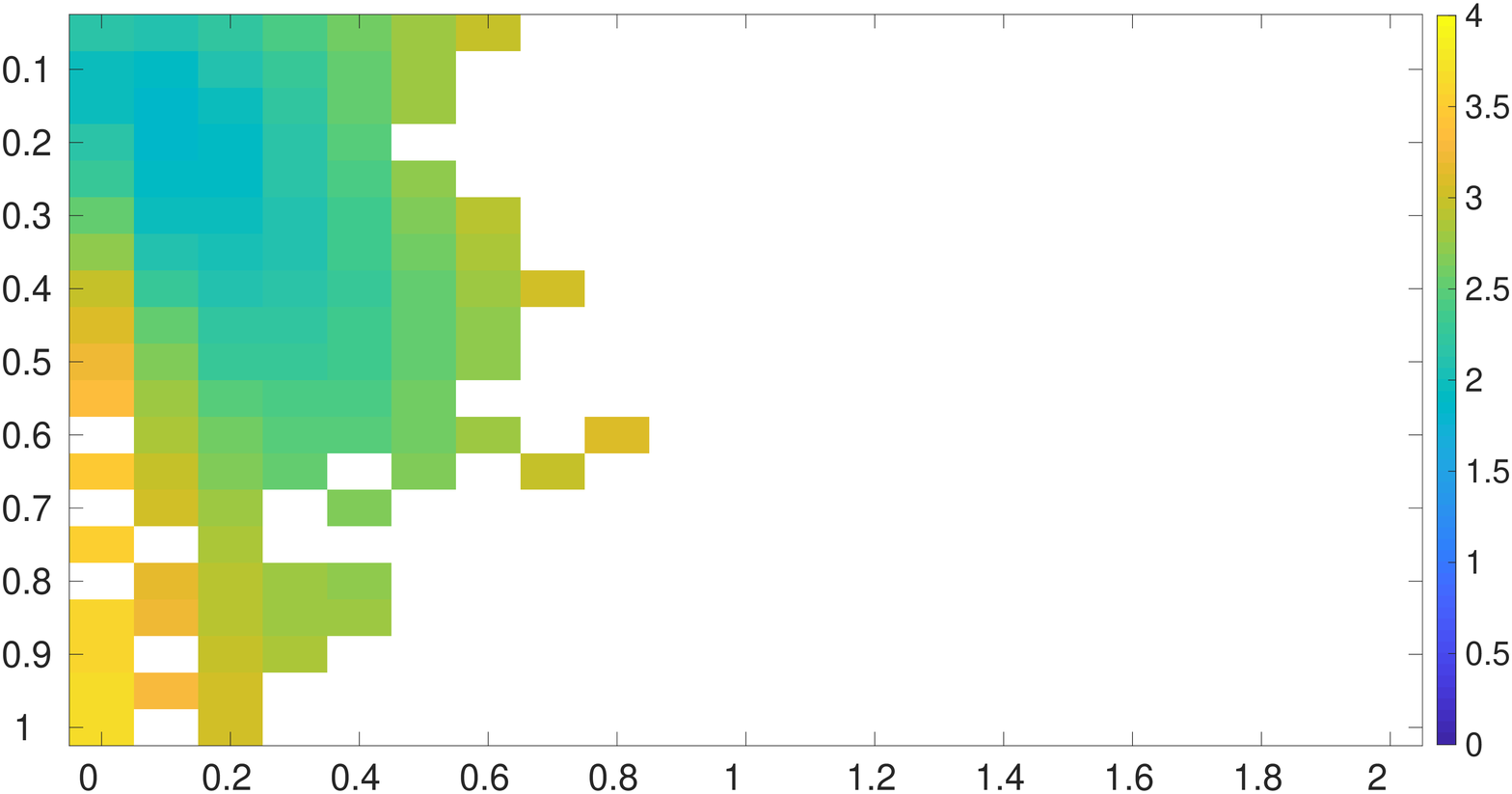}} \\ 
	\end{tabular}
	\caption{\label{fig:rmse_inf_loc_frequency} As in Figure \ref{fig:rmse_inf_loc_full_obs}, but for average RMSEs obtained by the grid search method in the half observation scenario, with the same ensemble sizes $N_e = 30$ yet different observation frequencies. For ease of comparison, the results of the half observation scenario ($\Delta n = 2$, $N^{freq} = 4$, $N_e = 30$) in Figure \ref{fig:rmse_inf_loc_half_quarter_obs} are re-plotted here.}
\end{figure} 

We investigate one more aspect, namely, the impact of observation frequency on the performance of the grid search method and the CHOP workflow. In line with this goal, we conduct three additional experiments, with the following  settings: $N_e = 30$, $\Delta n = 2$ (the half observation scenario), and $N^{freq} = 1, 2, 8$ respectively. 

Figure \ref{fig:rmse_inf_loc_frequency} shows the average RMSEs of the grid search method, when the inflation factor and the length scale take different values, and the observations arrive at different frequencies. For convenience of comparison, the results with $N^{freq} = 4$ ($N_e = 30$, $\Delta n = 2$) in Figure \ref{fig:rmse_inf_loc_half_quarter_obs} are also included into Figure \ref{fig:rmse_inf_loc_frequency}. It can be clearly seen that, as the observation frequency decreases (corresponding to increasing $N^{freq}$), the average RMSE tends to increase. Filter divergence remains a problem, but in this case, it appears that a lower observation frequency does not necessarily lead to a higher chance of filter divergence.

Following Tables \ref{tab:min_rmse_full_obs} and \ref{tab:min_rmse_half_quarter_obs}, Table \ref{tab:min_rmse_frequency} summarizes the minimum average RMSEs of the grid search method at different observation frequencies, their associated STDs and the optimal inflation factor and length scale. As observed in Table \ref{tab:min_rmse_frequency}, when the observation frequency decreases ($N^{freq}$ increases), the minimum average RMSE arises. In the meantime, the corresponding optimal length scale $\lambda_{min}$ tends to decline, while the optimal inflation factor $\delta_{min}$ remains unchanged. 

In terms of the performance of the CHOP workflow, one can observe again that its average RMSEs stay relatively close to the corresponding minimum values of the grid search method. On the other hand, no filter divergence is found in the repetition runs of the CHOP workflow. Altogether, the experiment results confirm that the CHOP workflow also performs reasonably well at different observation frequencies.    

\subsection{Experiments in a 1000-dimensional L96 system}
\begin{table*} [!t]
	\centering
	\caption{\label{tab:min_rmse_N1000} Performance comparison between the grid search method and the CHOP workflow in the 1000-dimensional L96 model.}
	\begin{tabular}{cccc}
		\hline
		\hline 
		\multicolumn{2}{c}{Grid search} & CHOP (SIF) & CHOP (MIF) \\
		\cline{1-4}
		Minimum average RMSE (mean $\pm$ STD) & $(\delta_{min},\lambda_{min})$ & Average RMSE (mean $\pm$ STD) & Average RMSE (mean $\pm$ STD) \\ 
		\hline  
		$2.7667 \pm 0.0099$ & $(0.10,0.05)$ & $3.4213 \pm  0.0552$ & $ 3.0264 \pm 0.0116$\\
		\hline             
		\hline
	\end{tabular}
\end{table*}   

\begin{figure} 
	\centering
	\includegraphics[scale=0.3]{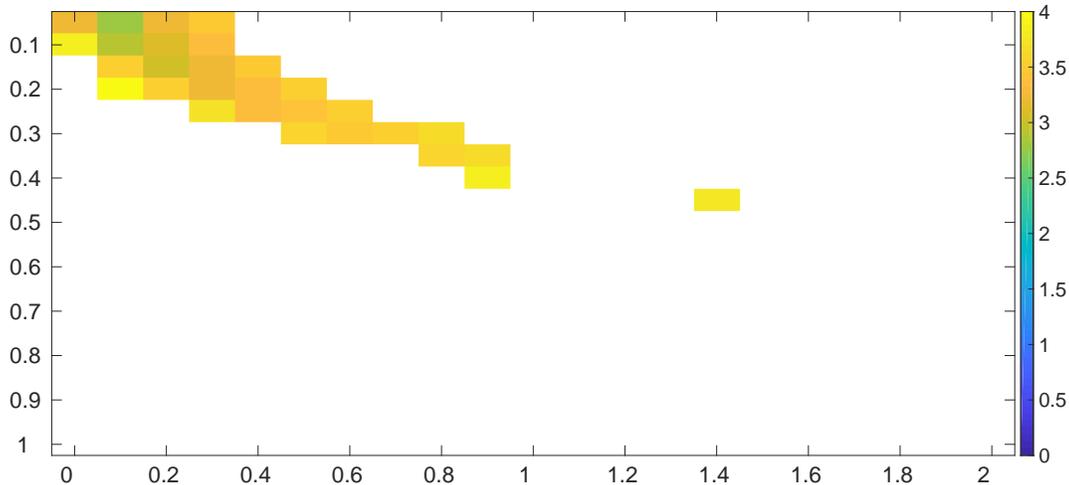}
	\caption{\label{fig:rmse_inf_loc_N1000} Average RMSEs obtained by the grid search method (applied to Eq. \ref{eq:SEnKF_inf_loc}) in the 1000-dimensional L96 model.}
\end{figure} 

In this subsection,  we conduct an additional experiment in a 1000-dimensional L96 model ($N_L = 1000$). The main purpose of the experiment is to demonstrate that the CHOP workflow can be used to tune a large number of hyper-parameters. This feature is a natural reflection of the capacity of the IES algorithm, which has been shown to work well in, e.g., large-scale reservoir data assimilation problems \citep{chen2013-levenberg,emerick2012ensemble,luo2015Iterative}. 

The experiment settings in this subsection is largely the same as those of the experiments with respect to the 40-dimensional L96 model. Therefore, for brevity, in the sequel we focus more on explaining the places where different experiment settings are adopted.    

Since the dimensionality is significantly increased, the grid search method becomes more time-consuming. To facilitate the investigation, we reduce the assimilation time window from 250 time units to 100 time units (corresponding to $2000$ integration steps), and the number of repetition runs of a given experiment from $20$ to $10$, while keeping the search ranges of the inflation factor and the length scale unchanged. In the meantime, we increase the ensemble size $N_e$ to $100$. The observation system is the same as that in Eq. \ref{eq:obs_system}, with the same observation-noise variance. The increment of model-variable index is set to $\Delta n = 4$ (quarter observation scenario), and the observations are collected every 4 integration steps ($N^{freq} = 4$). Given the purpose of the current experiment, no sensitivity study (e.g., with respect to $N_e$, $\Delta n$ and $N^{freq}$) is conducted.   

The base assimilation algorithm is the same as that in Eq. \ref{eq:SEnKF}, and we introduce both covariance inflation and localization to the base algorithm. We use the same localization scheme as in the 40-dimensional case (with the length scale $\lambda$ as a hyper-parameter), while considering two different ways of conducting covariance inflation. One inflation method is again the same as that in the 40-dimensional case, which applies a single inflation factor $\delta$ to all model state variables of the background ensemble. This leads to a reference algorithm identical to that in Eq. \ref{eq:SEnKF_inf_loc}, which contains two hyper-parameters, $\delta$ and $\lambda$, and the grid search method is then applied to find the optimal combination of $\delta$ and $\lambda$ for the reference algorithm. On the other hand, the CHOP workflow is employed to estimate an ensemble of $N_e$ hyper-parameter pairs $\left\{(\delta_j,\lambda_j) \right\}_{j=1}^{N_e}$. For distinction later, we call the application of the CHOP workflow to estimate the ensemble $\left\{(\delta_j,\lambda_j) \right\}_{j=1}^{N_e}$ the single-inflation-factor (SIF) method. 

The other inflation method introduces multiple inflation factors to the base algorithm. Specifically, each model state variable of the background ensemble $\mathcal{M}^b = \{\mathbf{m}_j^b\}_{j=1}^{N_e}$ receives its own inflation factor, in such a way that after inflation, the modified background ensemble  $\tilde{\mathcal{M}}^b \equiv \{\tilde{\mathbf{m}}_j^b\}_{j=1}^{N_e}$ has its member $\tilde{\mathbf{m}}_j^b$ in the form of $\tilde{\mathbf{m}}_j^b = \bar{\mathbf{m}}^b + \left( \boldsymbol{1}+\boldsymbol{\delta} \right) \circ \left( \mathbf{m}_j^b -  \bar{\mathbf{m}}^b \right)$, where $\boldsymbol{1}$ is a $N_L$-dimensional vector with all its elements equal to 1, $\boldsymbol{\delta} = \left[\delta_1,\delta_2,\dotsb,\delta_{N_L}\right]^T$ contains $N_L$ inflation factors, and $\circ$ stands for the Schur product operator. Replacing the SIF method in Eq. \ref{eq:SEnKF_inf_loc} by the multiple-factor one (while keeping the localization scheme unchanged), one obtains a new reference algorithm. 
{
	\begin{linenomath*}  
		\begin{IEEEeqnarray}{l} \label{eq:SEnKF_inf_loc_MIF}
			\mathbf{m}_j^a =  \tilde{\mathbf{m}}_j^b  +  \left\{ \mathbf{L}\left(\lambda \right) \circ \left[ \tilde{\mathbf{C}}_m \mathbf{H}^T \left( \mathbf{H} \tilde{\mathbf{C}}_m \mathbf{H}^T + \mathbf{C}_d  \right)^{-1} \right] \right\}  \left(\mathbf{d}_j^o - \mathbf{H} \tilde{\mathbf{m}}_j^b \right); \\
			\tilde{\mathbf{m}}_j^b  = \bar{\mathbf{m}}^b + \left( \boldsymbol{1}+\boldsymbol{\delta} \right) \circ \left( \mathbf{m}_j^b -  \bar{\mathbf{m}}^b \right),
		\end{IEEEeqnarray} 
	\end{linenomath*}
}   
where $\tilde{\mathbf{C}}_m$ is the sample covariance matrix with respected to the inflated ensemble $\tilde{\mathcal{M}}^b$. 

Due to the high dimensionality ($N_L = 1000$), it is computationally prohibitive to apply the grid search method to optimize the set of hyper-parameters in Eq. \ref{eq:SEnKF_inf_loc_MIF}. On the other hand, as will be shown later, it is still possible to apply the CHOP workflow to estimate an ensemble of hyper-parameters, denoted by $\left\{(\boldsymbol{\delta}_j,\lambda_j) \right\}_{j=1}^{N_e}$. Such a workflow is called the multiple-inflation-factor (MIF) method hereafter. 

With these said, in the sequel, we compare the performance of the grid search method applied to the reference algorithm in Eq. \ref{eq:SEnKF_inf_loc}, the CHOP workflow with the SIF method, and the CHOP workflow with the MIF method, respectively. 

Figure \ref{fig:rmse_inf_loc_N1000} shows the average RMSEs obtained by the grid search method with different combinations of $\delta$ and $\lambda$ values. Similar to what we have seen in the 40-dimensional L96 model, filter divergence arises in a large portion of the searched region of hyper-parameters. As reported in Table \ref{tab:min_rmse_N1000}, the minimum average RMSE of the grid search method is around $2.7667$, achieved at $\delta_{min} = 0.10$ and $\lambda_{min}=0.05$.  

For comparison, Table \ref{tab:min_rmse_N1000} also presents the average RMSEs of the CHOP workflow equipped with the SIF and MIF methods, respectively. Again, no filter divergence takes place in the CHOP workflow. Both the SIF and MIF methods result in RMSE values that stay relatively close to the minimum RMSE value of the grid search method. In comparison to the SIF method, however, the MIF exhibits better performance, largely due to a higher degree of freedom brought in by the larger number of inflation factors used in the assimilation algorithm.          

\subsection{Behavior of the IES algorithm}
\begin{figure} 
	\centering
	\includegraphics[scale=0.3]{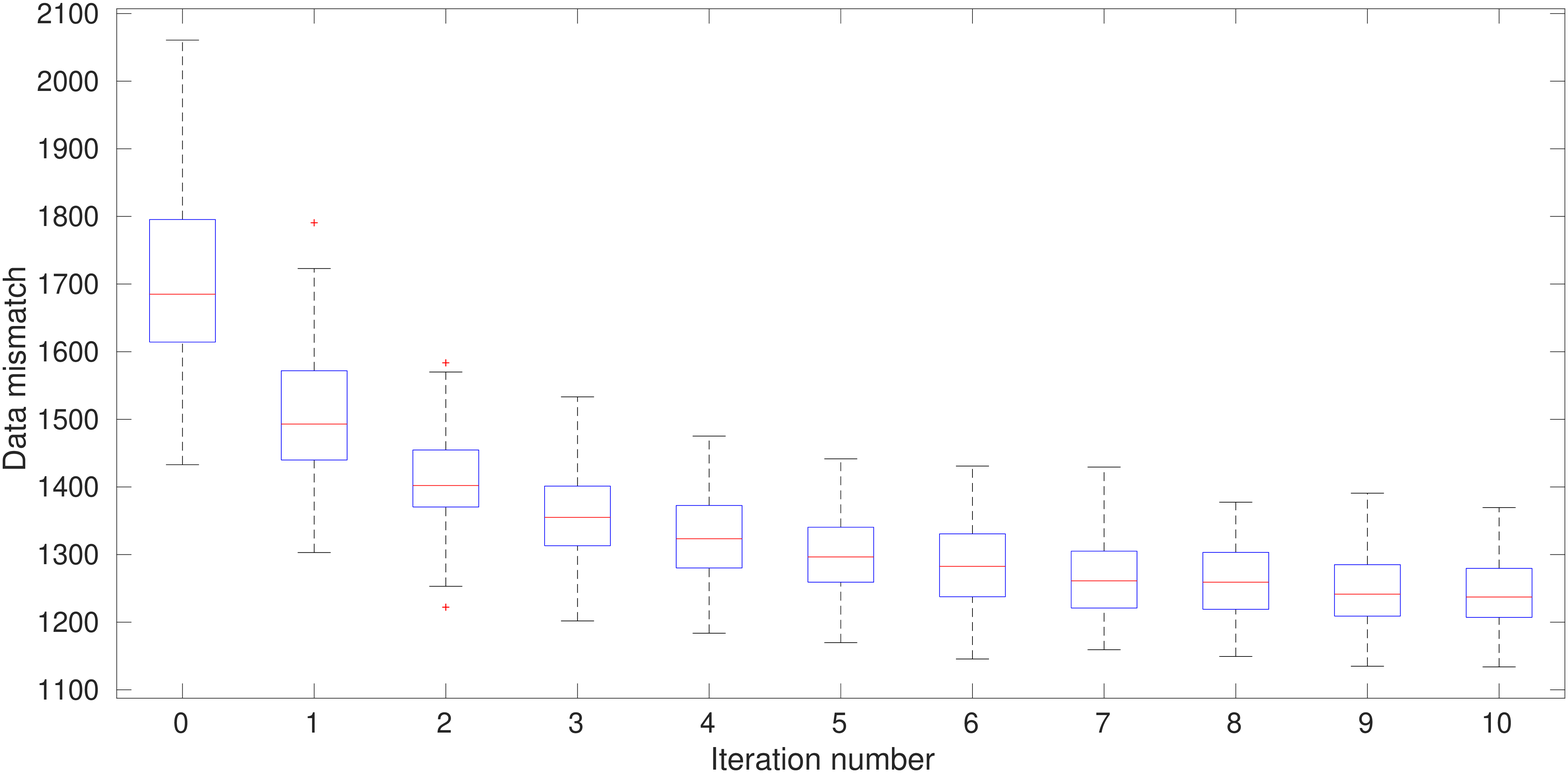}
	\caption{\label{fig:data_mismatch_N1000} Box plots of data mismatch at different iteration steps at one of the data assimilation cycles of the 1000-dimensional L96 model.}
\end{figure}

\begin{figure} 
	\centering
	\includegraphics[scale=0.3]{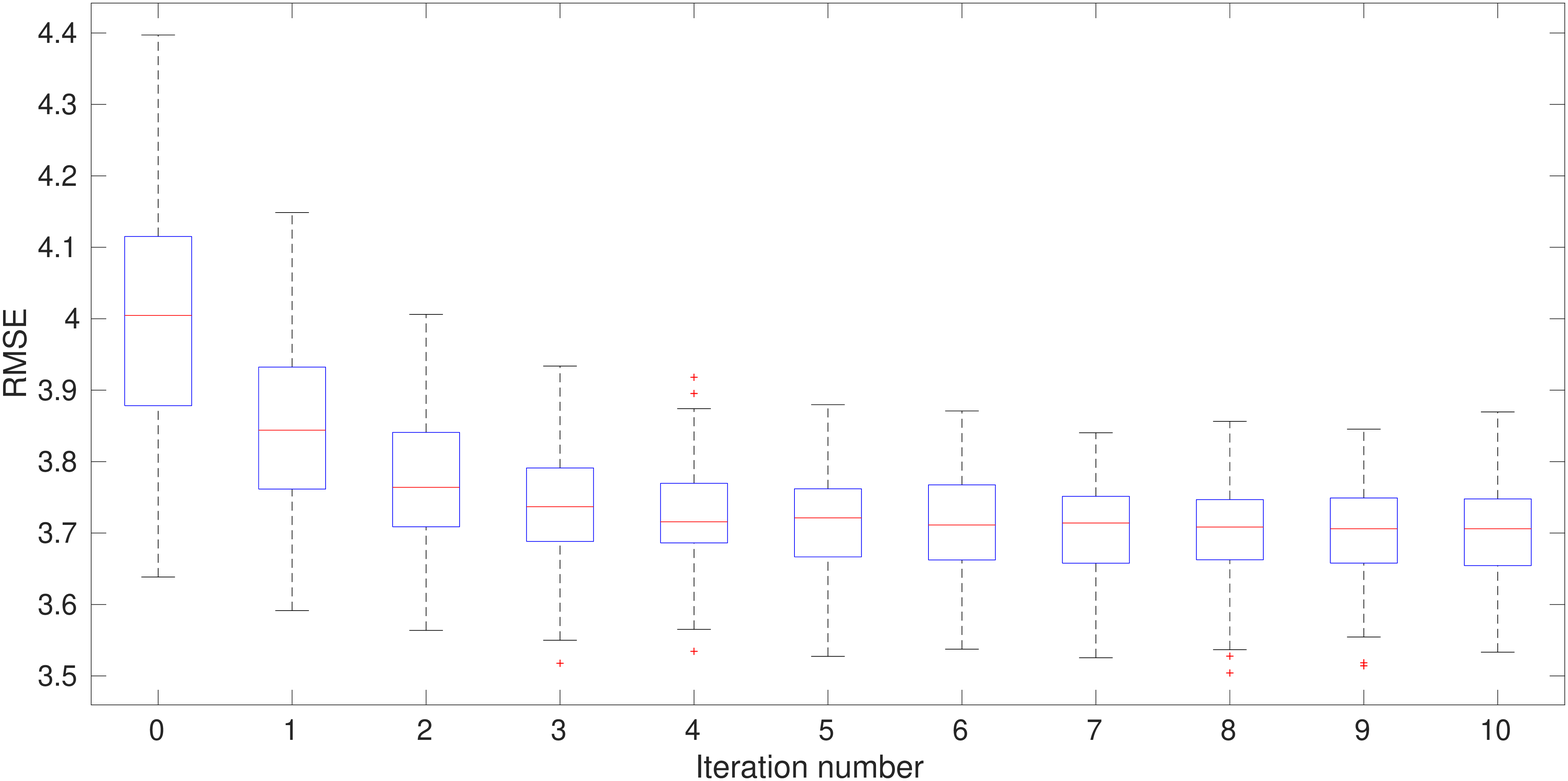}
	\caption{\label{fig:rmse_N1000} Box plots of RMSE at different iteration steps at one of the data assimilation cycles of the 1000-dimensional L96 model.}
\end{figure}

\begin{figure} 
	\centering
	\includegraphics[scale=0.3]{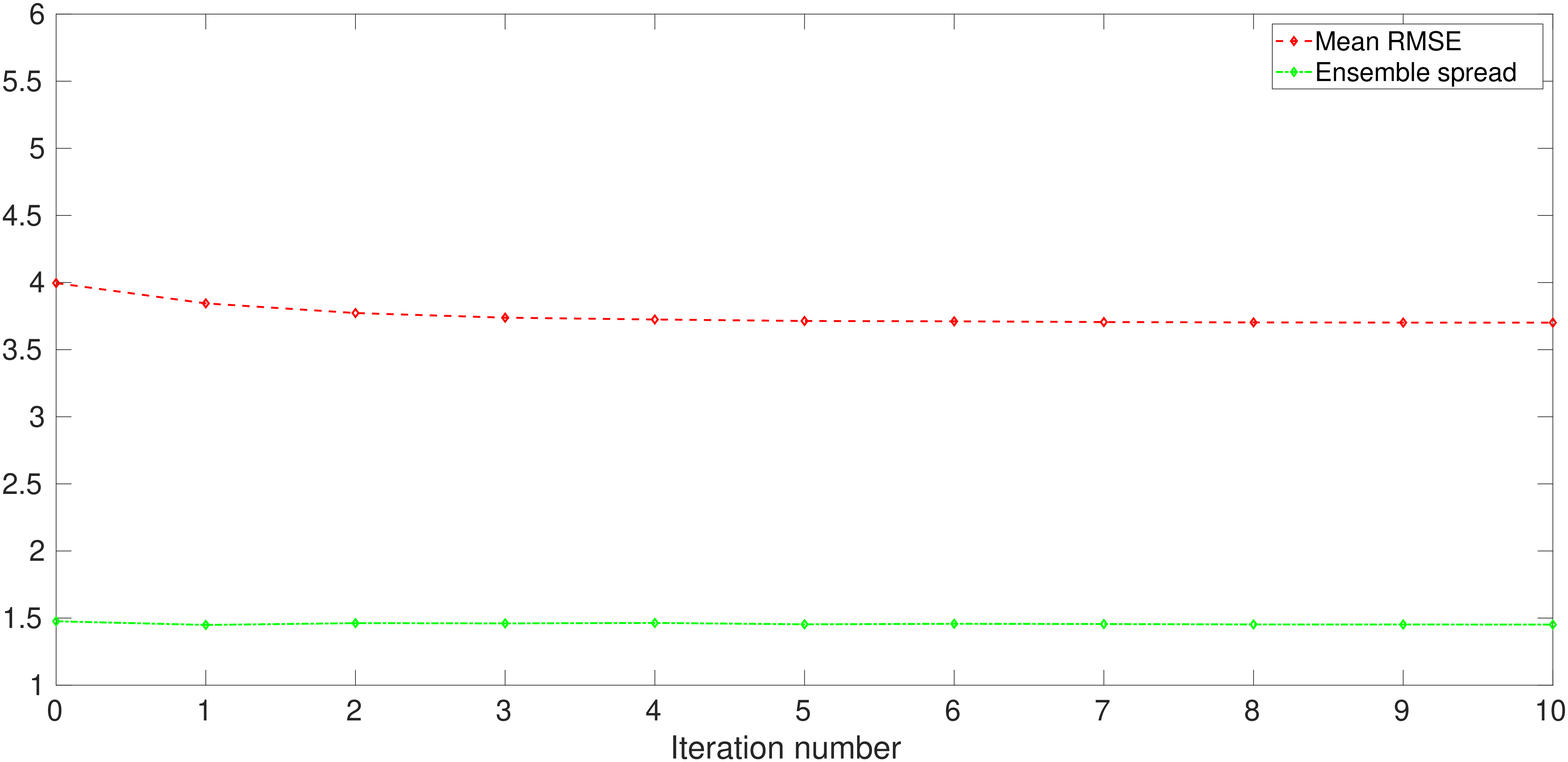}
	\caption{\label{fig:ensemble_spread_N1000} Mean RMSE (dashed red line) and ensemble spread (dash-dotted green line) versus iteration step, at one of the data assimilation cycles of the 1000-dimensional L96 model.}
\end{figure}

\renewcommand{\nScale}{0.15}
\begin{figure} 
	\centering
	\begin{tabular}{cc}
		\multicolumn{2}{c}{\subfloat[Histogram of the reference model state (truth)]{\includegraphics[scale=\nScale]{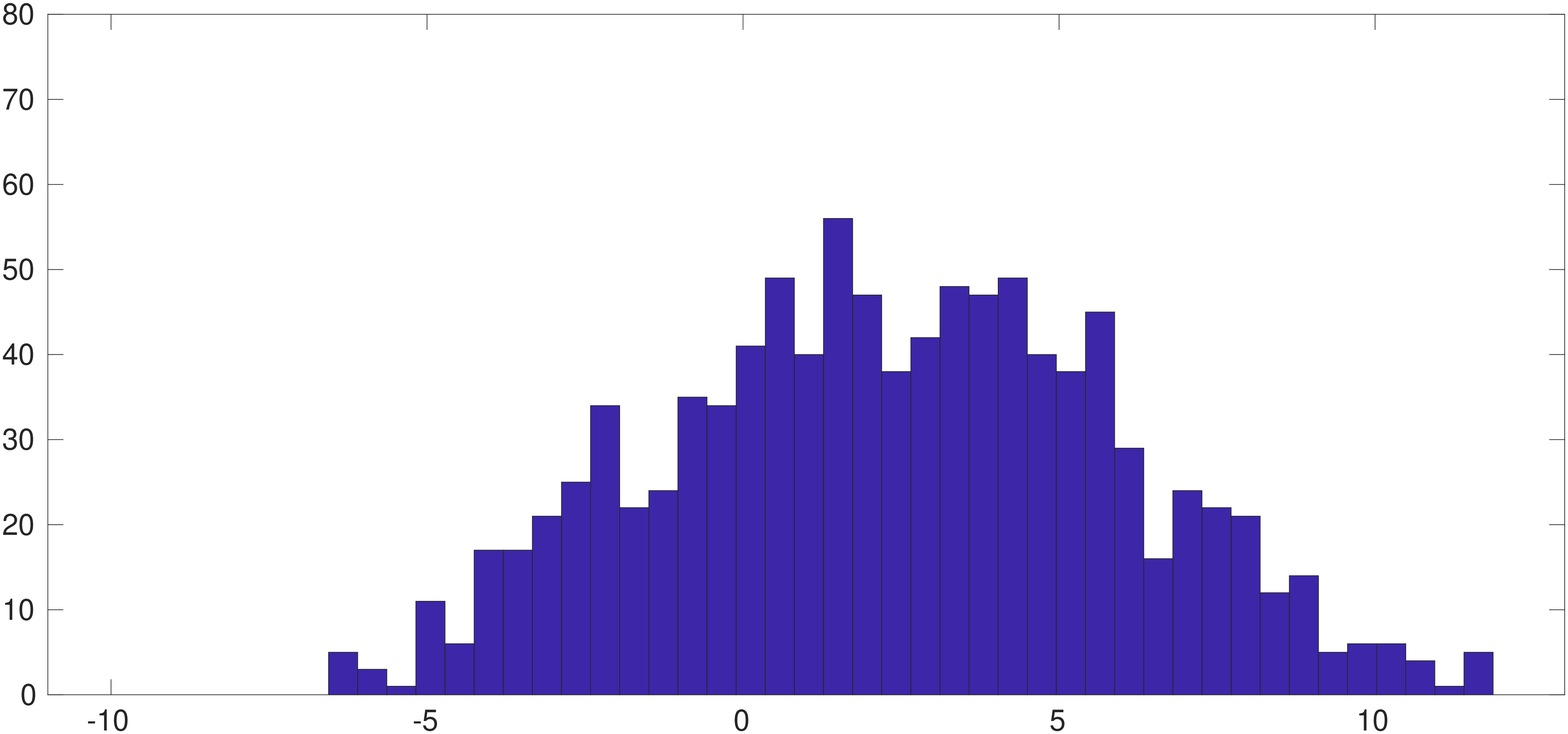}} }  \\
		\subfloat[Histogram of the background ensemble mean]{\includegraphics[scale=\nScale]{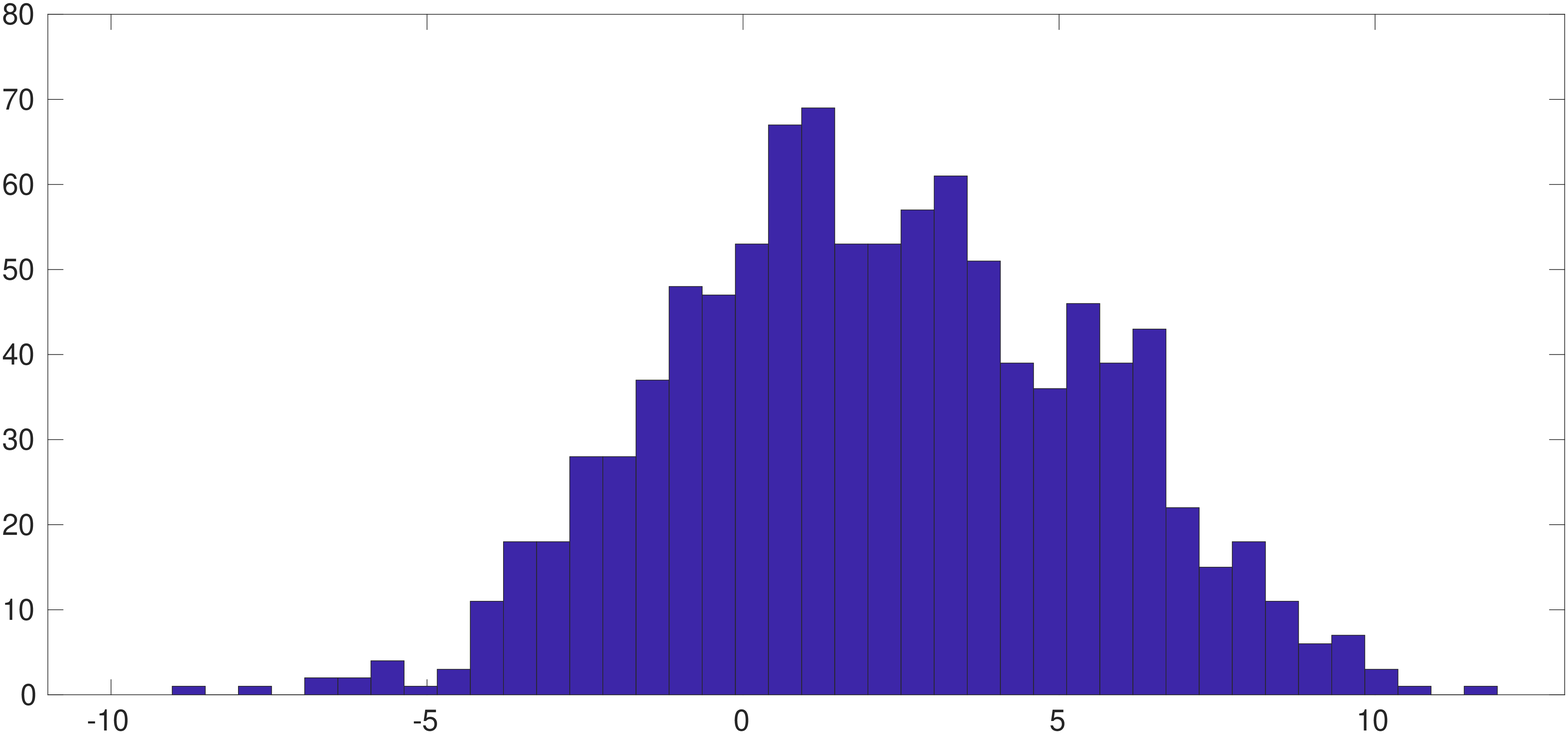}}
		&
		\subfloat[Histogram of the analysis ensemble mean]{\includegraphics[scale=\nScale]{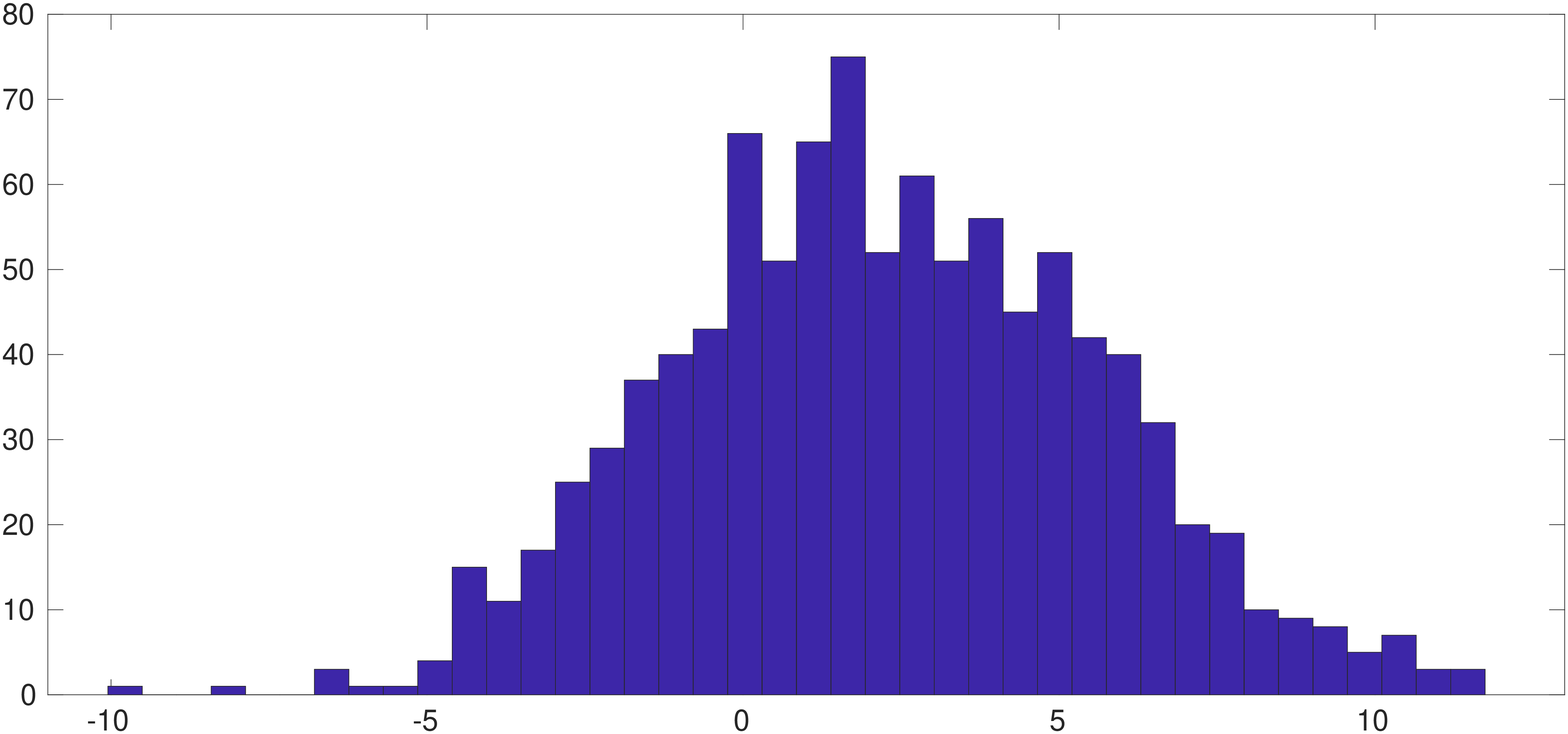}} \\ 
	\end{tabular}
	\caption{\label{fig:histogram_model_state} Histograms of (a) the reference model state (truth), (b) the background ensemble mean and (c) the analysis ensemble mean at one of the assimilation cycles in the 1000-dimensional L96 model. Both the reference model state and the background ensemble do not change over the IES iteration process, whereas the analysis ensemble is obtained by inserting the ensemble of estimated hyper-parameters at the last iteration step into the reference algorithm, Eq. \ref{eq:SEnKF_inf_loc_MIF}, of the MIF method.}
\end{figure} 

\renewcommand{\nScale}{0.15}
\begin{figure} 
	\centering
	\begin{tabular}{cc}
		\subfloat[Histogram of the initial ensemlbe of the inflation factor associated with model variable 1]{\includegraphics[scale=\nScale]{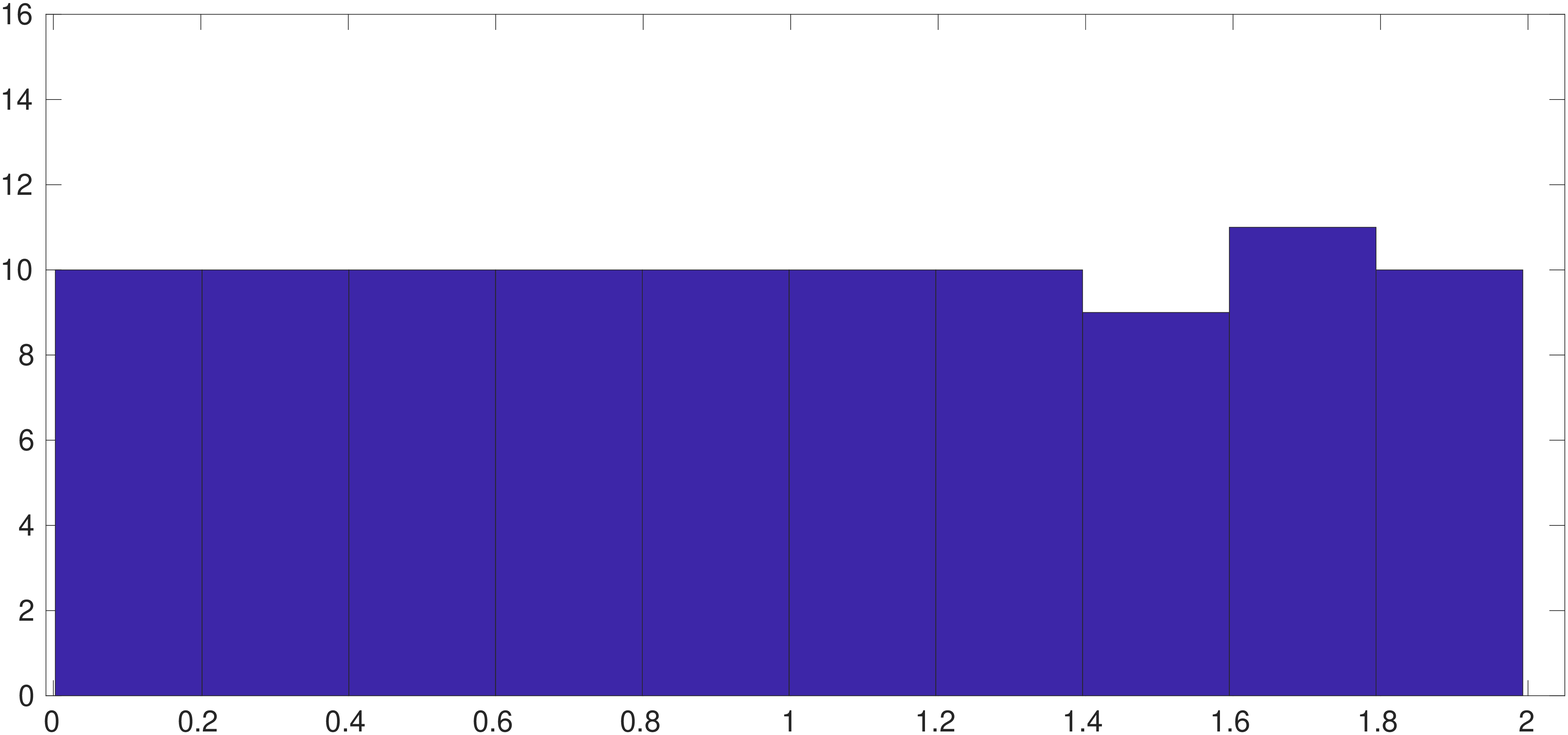}} 
		&
		\subfloat[Histogram of the final ensemlbe of the inflation factor associated with model variable 1]{\includegraphics[scale=\nScale]{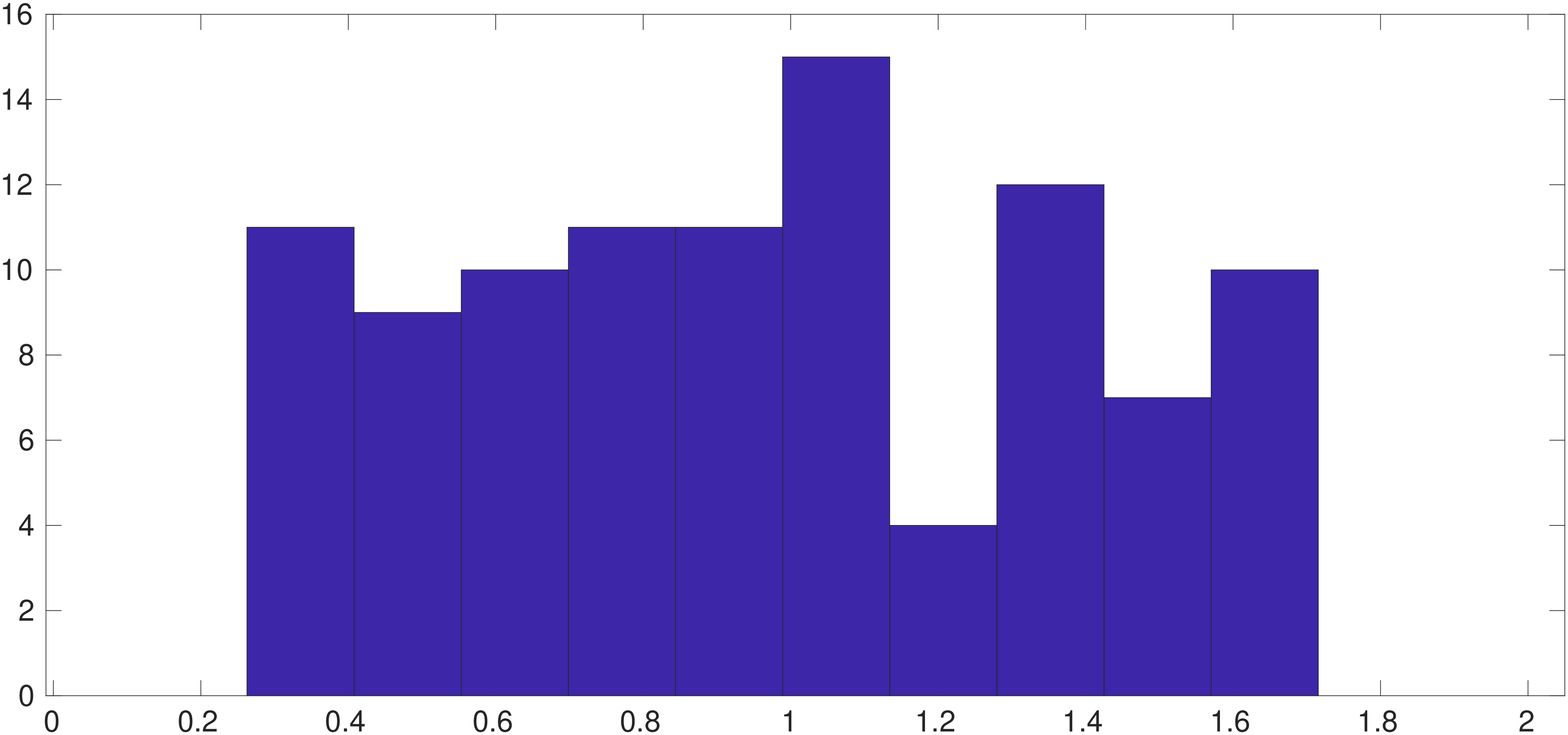}}  \\
		\subfloat[Histogram of the initial ensemlbe of the inflation factor associated with model variable 500]{\includegraphics[scale=\nScale]{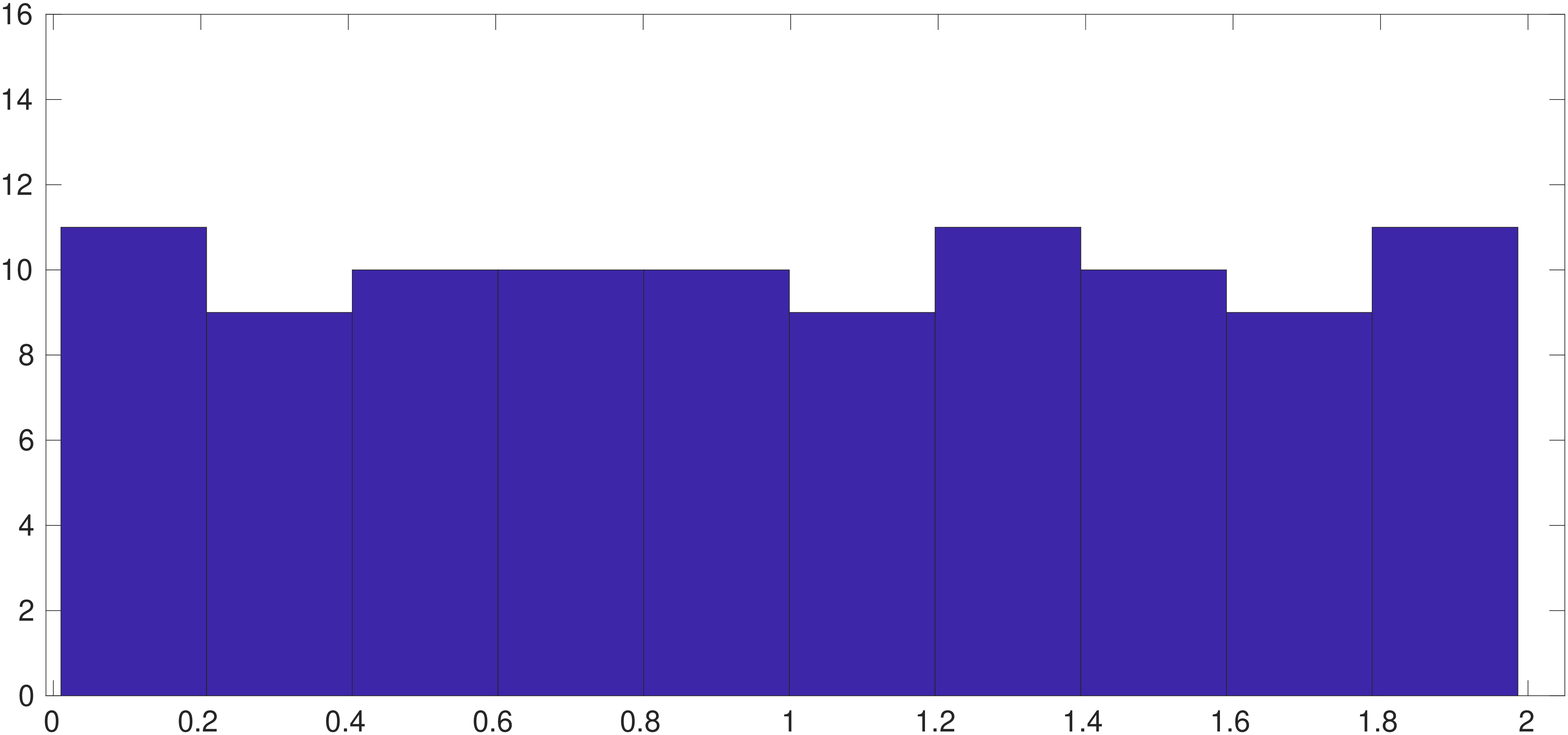}} 
		&
		\subfloat[Histogram of the final ensemlbe of the inflation factor associated with model variable 500]{\includegraphics[scale=\nScale]{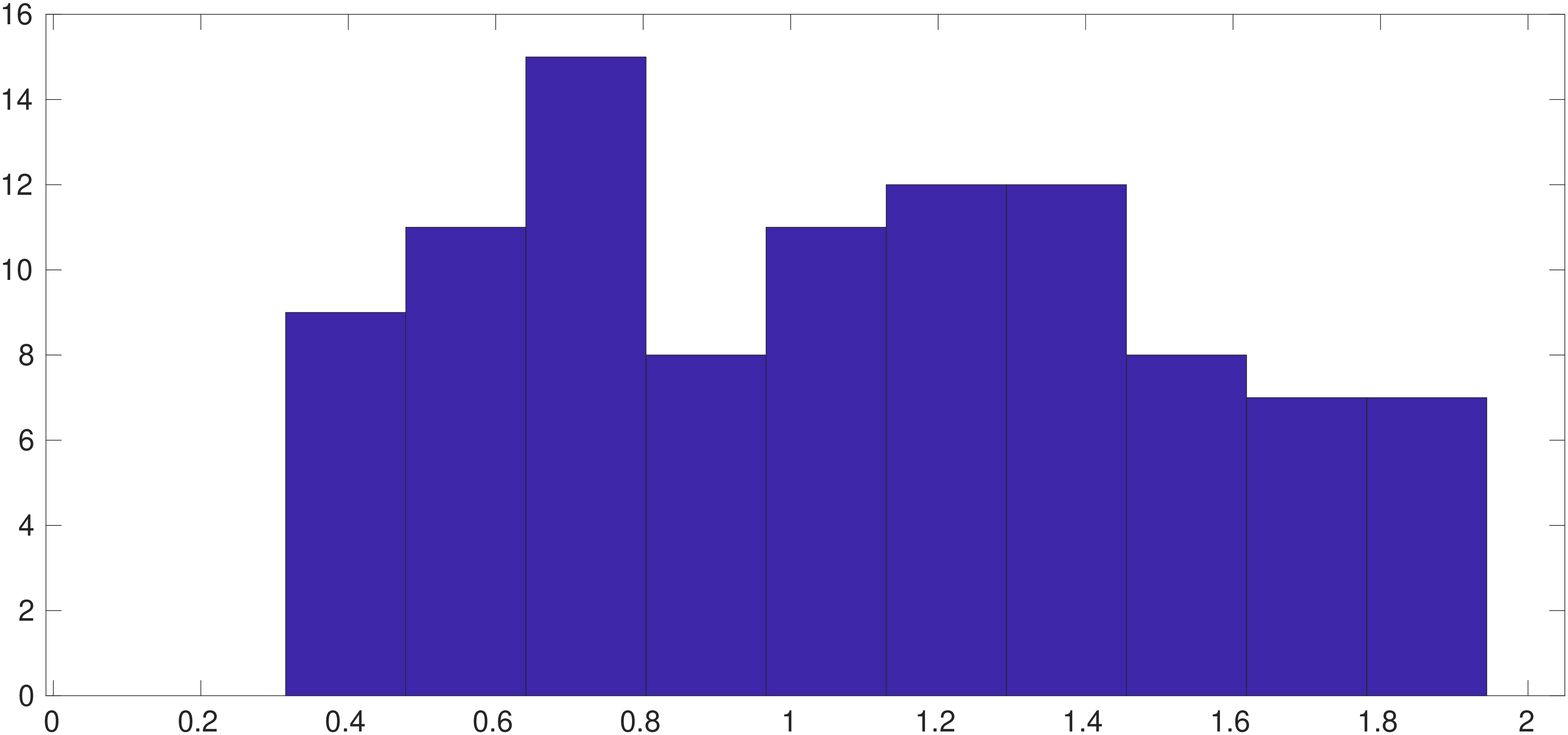}}  \\
		\subfloat[Histogram of the initial ensemlbe of the length scale]{\includegraphics[scale=\nScale]{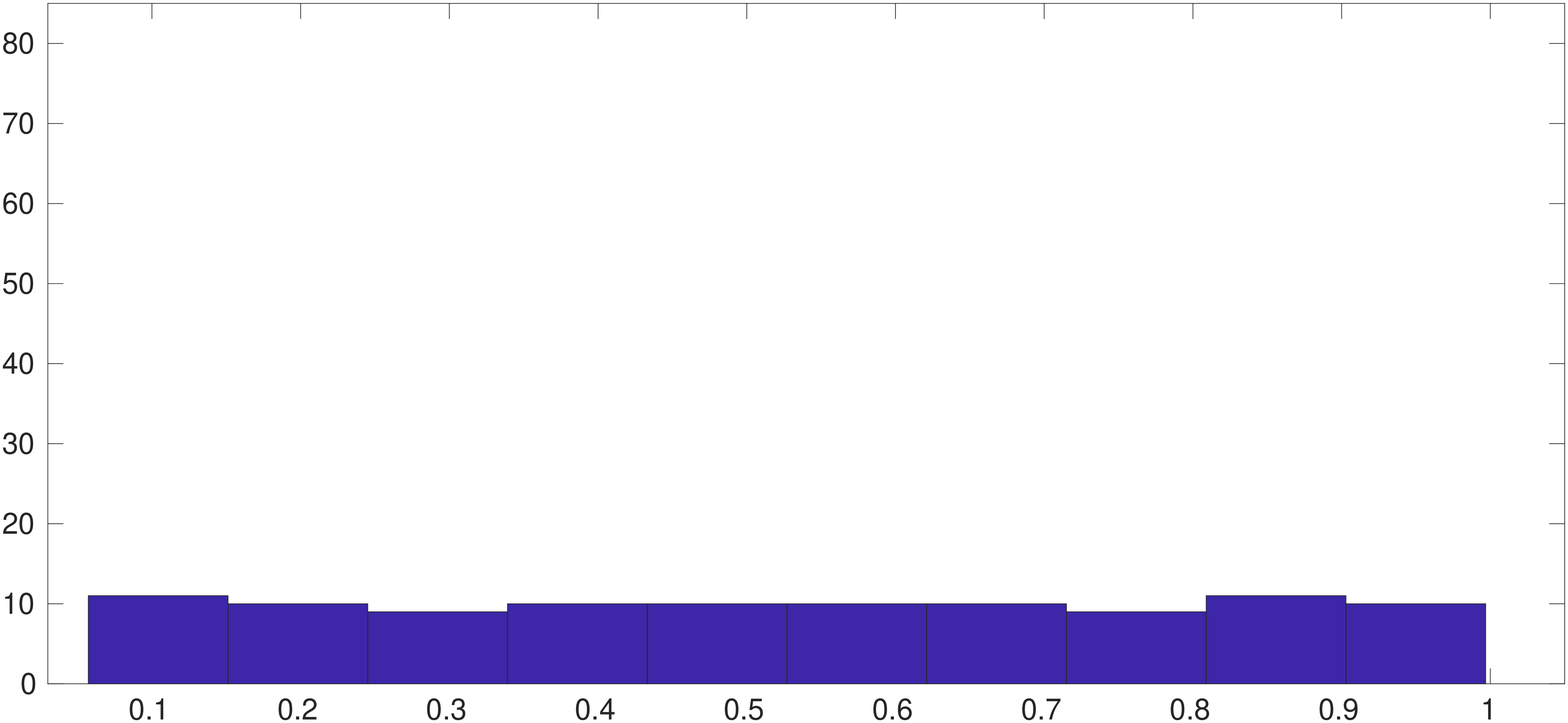}} 
		&
		\subfloat[Histogram of the final ensemlbe of the length scale]{\includegraphics[scale=\nScale]{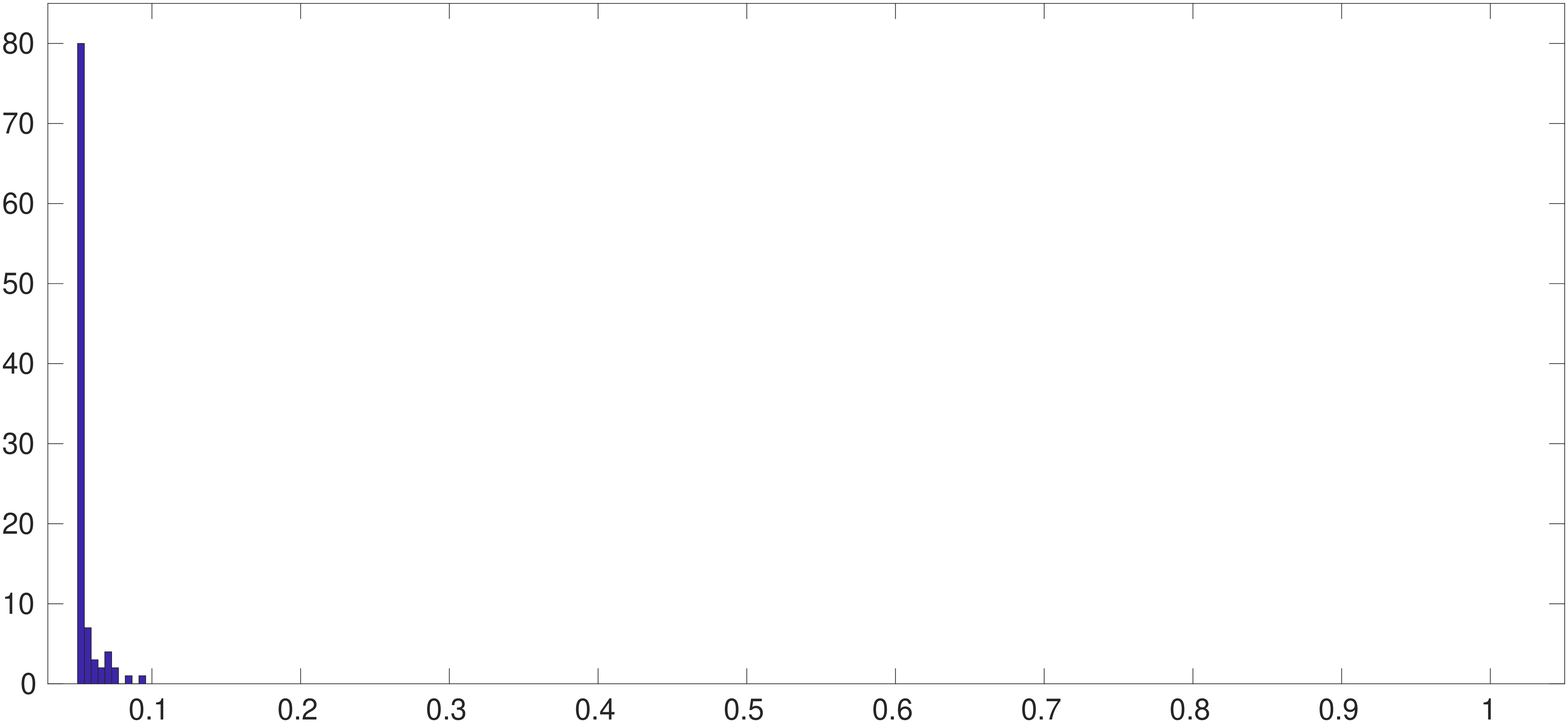}}  \\
	\end{tabular}
	\caption{\label{fig:histogram_hyper_parameters} Histograms of the initial (left) and final (right) ensembles, with respect to the inflation factors associated with model state variables 1 (top) and 500 (middle), and the length scale (bottom), respectively.}
\end{figure} 

Finally we take a glance at the behavior of the IES algorithm that underpins the CHOP workflow. We do this in the 1000-dimensional L96 model with the MIF method, to illustrate the efficacy of the IES algorithm in dealing with high-dimensional problems. Note that in the CHOP workflow, the IES is adopted to tune hyper-parameters at each assimilation cycle. For brevity, we only use one of the assimilation cycles for illustration.

Figures \ref{fig:data_mismatch_N1000} and \ref{fig:rmse_N1000} disclose the data mismatch and RMSE values at each iteration step, in the form of box plots. These values are obtained as follows: At each iteration step, we first insert the ensemble of hyper-parameters into the reference algorithm Eq. \ref{eq:SEnKF_inf_loc_MIF} of the MIF method, in such a way that each member of the background ensemble (of model state variables) is associated with a member of the ensemble of hyper-parameters. In this way, we obtain an ensemble of updated model state variables at each iteration step. The data mismatch and RMSE values are then calculated with respect to the ensemble of updated model state variables. Note that the ensemble of analysis state variables corresponds to the ensemble of updated model state variables at the last iteration step. Meanwhile, at iteration step 0, the data mismatch and RMSE values are computed based on the initial ensemble of hyper-parameters generated through the LHS scheme.   

In Figures \ref{fig:data_mismatch_N1000} and \ref{fig:rmse_N1000}, both the data mismatch and RMSE values tend to decrease as the iteration proceeds, while maintaining substantial ensemble varieties in the box plots (indicating that ensemble collapse does not take place). The IES converges relatively fast, moving into the vicinity of a certain local minimum after only several iteration steps, which is a behavior also noticed in other studies \citep{chen2013-levenberg,emerick2012ensemble,luo2015Iterative}. 

Corresponding to Figures \ref{fig:data_mismatch_N1000} and \ref{fig:rmse_N1000}, Figure \ref{fig:ensemble_spread_N1000} presents the values of mean RMSE and ensemble spread at each iteration step. Here, a mean RMSE is the average of the RMSEs over ensemble members of the updated model state variables (i.e., the average of the box-plot values) at a given iteration step, whereas ensemble spread is evaluated according to Eq. \ref{eq:ensemble_spread}. In consistency with Figure \ref{fig:rmse_N1000}, the mean RMSE and the ensemble spread tend to decrease along with the iterations. The overall change of ensemble spread from the beginning to the end of the iteration process appears to be less significant than that of the mean RMSE. In fact, the final ensemble spread appears to stay close to the initial value, which also suggests that ensemble collapse does not appear to be a problem. On the other hand, there are substantial gaps between the values of mean RMSE and ensemble spread at all iteration steps, which means that ensemble spread does not match the estimation errors of the updated model state variables. This tendency of under-estimation seems to be largely related to the fact that the ensemble spread at the beginning of the iteration is already considerably smaller than the mean RMSE, which could be due to the insufficient ensemble spread in the background ensemble, or the initial ensemble of hyper-parameters, or both. 

Figure \ref{fig:histogram_model_state} shows the histograms with respect to the reference model state variables (the truth), the background-ensemble mean, and the analysis-ensemble mean, respectively. It is clear that neither the histogram of the background-ensemble mean, nor that of the analysis-ensemble mean, resemble the histogram of the truth well, suggesting that there are substantial estimation errors in the estimated model state variables. 

On the other hand, the results with respect to the estimated hyper-parameters appears to be more interesting. For illustration, Figure \ref{fig:histogram_hyper_parameters} plots the histograms of the initial (left) and final (right) ensembles of the inflation factors associated with model state variable 1 (top) and 500 (middle), and the histograms of the  initial and final ensembles of the length scale (bottom). Since we use LHS to generate the initial ensemble, it can be observed that the histograms with respect to three initial ensembles of hyper-parameters roughly follow certain uniform distributions. Through the iteration process of the IES algorithm, the shapes and supports of the histograms are modified. This is particularly noticeable for the estimated values of length scale in the final ensemble (Figure \ref{fig:histogram_hyper_parameters}(f)). Initially, the range of the length scale in the initial ensemble is $[0.05,1]$, at the end of the iteration, around 80\% of the values of estimated length scale locate at 0.05 (which is the optimal value found by the grid search method), while the rest of the estimated values are less than 0.1. On the other hand, for the estimated inflation factors, one may notice that their values are less concentrated than the length scale. In comparison to the initial ensembles of the inflation factors, their final ensembles receive somewhat narrower supports, but still maintain sufficient spreads, in consistency with the results in Figure \ref{fig:ensemble_spread_N1000}. The values of estimated inflation factors are substantially larger than the optimal inflation factor (0.10) found by the grid search method. The main reason behind this is that the original EnKF updates model state variables only once, whereas the CHOP workflow does the update multiple times, each time with a smaller step size (hence larger inflation factors).


\section{Discussion and conclusion}  
This study aims to develop a Continuous Hyper-parameter Optimization (CHOP) workflow that helps to tune hyper-parameters in ensemble data assimilation algorithms. The main idea is to treat a data assimilation algorithm with certain hyper-parameters as a parametric mapping that transforms an ensemble of initial model state variables and/or parameters to a corresponding ensemble of updated quantities, which in turn are related to the predicted observations through the observation operator. 

Following this perspective, the hyper-parameters can be tuned in such a way that the corresponding updated model state variables and/or parameters result in lower data mismatch than their initial values. In doing so, the CHOP problem is recast as a parameter estimation problem. We adopt an iterative ensemble smoother (IES) to solve the CHOP problem, as its derive-free nature allows one to implement the algorithm without explicitly knowing the relevant gradients. To mitigate the adverse effects of using a relatively small ensemble size in the IES, we also equip the IES with a correlation-based adaptive localization scheme, which helps to handle the issue that hyper-parameters may not possess physical locations needed for distance-based localization schemes. 

We investigate the performance of the CHOP workflow in the Lorentz 96 (L96) model with two different dimensions. Experiments in the 40-dimensional L96 model aim to inspect the impacts of a few factors on the performance of the CHOP workflow, whereas those in the 1000-dimensional L96 model focus on demonstrating the capacity of the CHOP workflow to deal with a high-dimensional set of hyper-parameters, which may not be computationally feasible for the grid search method. Such a capacity would help enable the developments of more sophisticated auxiliary techniques (e.g., inflation or localization) that introduce a large number of hyper-parameters to an assimilation algorithm for further performance improvements. 

In most of the experiments, the CHOP workflow is able to achieve reasonably good performance, which is relatively close to the best performance obtained by the grid search method (an unverifiable case occurs in the experiments with respect to the multiple-inflation-factor method in the 1000-dimensional L96 model, where we are not able to adopt the grid search method due to its prohibitively expensive cost). Meanwhile, unlike the grid search method, the optimality criterion in the CHOP workflow is based on data mismatch between real and predicted observations, which is realistic and can be implemented in practical data assimilation problems.  

So far, we have only implemented the CHOP workflow in the ensemble Kalman filter (EnKF) with perturbed observations. Given the varieties of different assimilation algorithms (some of them may not even be ensemble-based), the way of implementing a CHOP workflow may have to adapt to the particular assimilation algorithm  in choice, which is an issue to be further studied in the future. On the other hand, though, we expect that the notion of treating an assimilation algorithm with hyper-parameters as a parametric mapping may still be valid. As such, it appears sensible that one converts a generic assimilation problem (being state estimation, parameter estimation or both) with hyper-parameters into a parameter estimation problem, and solve it through a certain iterative assimilation algorithm.  

\backmatter





\bmhead{Acknowledgments}
X. Luo acknowledges financial supports from the NORCE research project “Assimilating 4D Seismic Data: Big Data Into Big Models” which is funded by industry partners, Equinor Energy AS, Lundin Energy Norway AS, Repsol Norge AS, Shell Global Solutions International B.V., TotalEnergies EP Norge AS, and Wintershall Dea Norge AS, as well as the Research Council of Norway (project number: 295002).

C. Xia acknowledges financial supports from the National Nature Science Foundation of China (Grant No. 42002247) and  the Nature Science Foundation of Guangdong Province, China (Grant No. 
2020A1515111054).

\section*{Declarations}
Data will be made available upon request.

\bibliography{references}


\end{document}